\documentclass[pra,twocolumn,showpacs,preprintnumbers]{revtex4}

\usepackage{graphicx}
\usepackage{bm}
\usepackage{amsthm}
\usepackage{amsfonts, graphicx, amsmath}

\DeclareGraphicsExtensions{.eps,.ps,.eps.gz,.ps.gz,.eps.Z}

\newcommand{\ra}{\rangle}
\newcommand{\la}{\langle}

\newcommand{\be}{\begin{equation}}
\newcommand{\ee}{\end{equation}}
\newcommand{\ber}{\begin{eqnarray}}
\newcommand{\eer}{\end{eqnarray}}

\begin{document}

\title{Non-Markovian dynamics of a qubit coupled to an Ising spin bath}
\author{Hari Krovi$^{(1)}$, Ognyan Oreshkov$^{(2)}$, Mikhail Ryazanov$^{(3)}$,
    Daniel A. Lidar$^{(1,2,3)}$
  }
\affiliation{$^{(1)}$Department of Electrical Engineering, $^{(2)}$Department
  of Physics, $^{(3)}$Department of Chemistry, University of Southern California, \\
Los Angeles, California 90089, USA}

\begin{abstract}
We study the analytically solvable Ising model of a single qubit system
coupled to a spin bath. The purpose of this study is to analyze and
elucidate the performance of Markovian and non-Markovian master equations
describing the dynamics of the system qubit, in comparison to the exact
solution. We find that the time-convolutionless master equation performs
particularly well up to fourth order in the system-bath coupling constant,
in comparison to the Nakajima-Zwanzig master equation. Markovian approaches
fare poorly due to the infinite bath correlation time in this model. A
recently proposed post-Markovian master equation performs comparably to the
time-convolutionless master equation for a properly chosen memory kernel,
and outperforms all the approximation methods considered here at long times.
Our findings shed light on the applicability of master equations to the
description of reduced system dynamics in the presence of spin-baths.
\end{abstract}

\maketitle

\section{Introduction}

A major conceptual as well as technical difficulty in the practical
implementation of quantum information processing and quantum control schemes
is the unavoidable interaction of quantum systems with their environment.
This interaction can destroy quantum superpositions and lead to an
irreversible loss of information, a process generally known as decoherence.
Understanding the dynamics of open quantum systems is therefore of
considerable importance. The Schr\"{o}dinger equation, which describes the
evolution of closed systems, is generally inapplicable to open systems,
unless one includes the environment in the description. This is, however,
generally difficult, due to the large number of environment degrees of
freedom. An alternative is to develop a description for the evolution of
only the subsystem of interest. A multitude of different approaches have
been developed in this direction, exact as well as approximate \cite%
{Alicki:87,Breuer:book}. Typically the exact approaches are of limited
practical usefulness as they are either phenomenological or involve
complicated integro-differential equations. The various approximations lead
to regions of validity that have some overlap. Such techniques have been
studied for many different models, but their performance in general, is not
fully understood.

In this work we consider an exactly solvable model of a single qubit (spin $%
1/2$ particle) coupled to an environment of qubits. We are motivated by the
physical importance of such spin bath models \cite{Prokofev:00} in the
description of decoherence in solid state quantum information processors,
such as systems based on the nuclear spin of donors in semiconductors \cite%
{Kane:98,Vrijen:00}, or on the electron spin in quantum dots \cite{Loss:98}.
Rather than trying to accurately model decoherence due to the spin bath in
such systems (as in, e.g., Refs. \cite{sousa:115322,Witzel:06}), our goal in this work
is to compare the performance of different master equations which have been
proposed in the literature. Because the model we consider is exactly
solvable, we are able to accurately assess the performance of the
approximation techniques that we study. In particular, we study the
Born-Markov and Born master equations, and the perturbation expansions of
the Nakajima-Zwanzig (NZ) \cite{Nakajima:58,Zwanzig:60a} and the
time-convolutionless (TCL) master equations \cite{Shibata:77,ShiAri80} up to
fourth order in the coupling constant. We also study the post-Markovian (PM)
master equation proposed in \cite{ShabaniLidar:05}.

The dynamics of the system qubit in the model we study is highly
non-Markovian and hence we do not expect the traditional Markovian master
equations commonly used, e.g., in quantum optics \cite{Carmichael:book} and
nuclear magnetic resonance \cite{Slichter:book}, to be accurate. This is
typical of spin-baths, and was noted, e.g., by Breuer et al. \cite{BBP04}.
The work by Breuer et al. (as well as by other authors in a number of
subsequent publications \cite%
{Palumbo:06,Burgarth:06,Hamdouni:06,Yuan:07,Camalet:07,Jing:07}) is
conceptually close to ours in that in both cases an analytically solvable
spin-bath model is considered and the analytical solution for the open
system dynamics is compared to approximations. However, there are also
important differences, namely, in Ref. \cite{BBP04} a so-called spin-star
system was studied, where the system spin has equal couplings to all the
bath spins, and these are of the XY exchange-type. In contrast, in our model
the system spin interacts via Ising couplings with the bath spins, and we
allow for arbitrary coupling constants. As a result there are also important
differences in the dynamics. For example, unlike the model in Ref. \cite%
{BBP04}, for our model we find that the odd order terms in the perturbation
expansions of Nakajima-Zwanzig and time-convolutionless master equations are
non-vanishing. This reflects the fact that there is a coupling between the $%
x $ and $y$ components of the Bloch vector which is absent in \cite{BBP04}.
In view of the non-Markovian behavior of our model, we also discuss the
relation between a representation of the analytical solution of our model in
terms of completely positive maps, and the Markovian limit obtained via a
coarse-graining method introduced in \cite{Lidar:CP01}, and the performance
of the post-Markovian master equation \cite{ShabaniLidar:05}.

This paper is organized as follows. In Sec II, we present the model, derive
the exact solution and discuss its behavior in the limit of small times and
large number of bath spins, and in the cases of discontinuous spectral
density co-domain and alternating sign of the system-bath coupling constants. In Sec. III, we consider second order
approximation methods such as the Born-Markov and Born master equations, and
a coarse-graining approach to the Markovian semigroup master equation. Then
we derive solutions to higher order corrections obtained from the
Nakajima-Zwanzig and time-convolutionless projection techniques as well as
derive the optimal approximation achievable through the post-Markovian
master equation. In Sec. IV, we compare these solutions for various
parameter values in the model and plot the results. Finally in Sec. V, we
present our conclusions.

\section{Exact dynamics}

\subsection{The model}

We consider a single spin-$\frac{1}{2}$ system (i.e., a qubit with a
two-dimensional Hilbert space $\mathcal{H}_{S}$) interacting with a bath of $%
N$ spin-$\frac{1}{2}$ particles (described by an $N$-fold tensor product of
two-dimensional Hilbert spaces denoted $\mathcal{H}_{B}$). We model the
interaction between the system qubit and the bath by the Ising Hamiltonian 
\begin{equation}
H_{I}^{\prime }=\alpha \sigma ^{z}\otimes \sum_{n=1}^{N}g_{n}\sigma _{n}^{z},
\label{eq:HI}
\end{equation}%
where $g_{n}$ are dimensionless real-valued coupling constants in the
interval $[-1,1]$, and $\alpha >0$ is a parameter having the dimension of
frequency (we work in units in which $\hbar =1$), which describes the
coupling strength and will be used below in conjunction with time ($\alpha t$%
) for perturbation expansions. The system and bath Hamiltonians are 
\begin{equation}
H_{S}=\frac{1}{2}\omega _{0}\sigma ^{z}  \label{SystemHam}
\end{equation}%
and 
\begin{equation}
H_{B}=\sum_{n=1}^{N}\frac{1}{2}\Omega _{n}\sigma _{n}^{z}.
\end{equation}%
For definiteness, we restrict the frequencies $\omega _{0}$ and $\Omega _{n}$
to the interval $[-1,1]$, in inverse time units. Even though the units of
time can be arbitrary, by doing so we do not lose generality, since we will
be working in the interaction picture where only the frequencies $\Omega _{n}
$ appear in relation to the state of the bath [Eq.~(\ref{eq:rhoB0})]. Since
the ratios of these frequencies and the temperature of the bath occur in the
equations, only their values relative to the temperature are of interest.
Therefore, henceforth we will omit the units of frequency and temperature
and will treat these quantities as dimensionless.

The interaction picture is defined as the transformation of any operator 
\begin{equation}
A\mapsto A(t)=\exp (iH_{0}t)A\exp (-iH_{0}t),
\end{equation}%
where $H_{0}=H_{S}+H_{B}$. The interaction Hamiltonian $H_{I}$ chosen here
is invariant under this transformation since it commutes with $H_{0}$. [Note
that in the next subsection, to simplify our calculations we redefine $H_{S}$
and $H_{I}^{\prime }$ (whence $H_{I}^{\prime }$ becomes $H_{I}$), but this
does not alter the present analysis.] All the quantities discussed in the
rest of this article are assumed to be in the interaction picture.

The dynamics can be described using the superoperator notation for the
Liouville operator 
\begin{equation}
\mathcal{L}\rho (t)\equiv -i[H_{I}^{\prime },\rho (t)],
\end{equation}%
where $\rho (t)$ is the density matrix for the total system in the Hilbert
space $\mathcal{H}_{S}\otimes \mathcal{H}_{B}$. The dynamics is governed by
the von Neumann equation 
\begin{equation}
\frac{d}{dt}\rho (t)=\alpha \mathcal{L}\rho (t)
\end{equation}%
and the formal solution of this equation can be written as follows: 
\begin{equation}
\rho (t)=\exp (\alpha \mathcal{L}t)\rho (0).  \label{VNeqnSoln}
\end{equation}%
The state of the system is given by the reduced density operator 
\begin{equation}
\rho _{S}(t)=\mathrm{Tr}_{B}\{\rho (t)\},
\end{equation}%
where $\mathrm{Tr}_{B}$ denotes a partial trace taken over the bath Hilbert
space $\mathcal{H}_{B}$. This can also be written in terms of the Bloch
sphere vector 
\begin{equation}
\vec{v}(t)=%
\begin{pmatrix}
v_{x}(t) \\ 
v_{y}(t) \\ 
v_{z}(t)%
\end{pmatrix}%
=\mathrm{Tr}\{\vec{\sigma}\rho _{S}(t)\},
\end{equation}%
where $\vec{\sigma}\equiv (\sigma ^{x},\sigma ^{y},\sigma ^{z})$ is the
vector of Pauli matrices. In the basis of $\sigma ^{z}$ eigenstates this is
equivalent to 
\begin{eqnarray}
\rho _{S}(t) &=&\frac{1}{2}(I+\vec{v}\cdot \vec{\sigma})  \notag \\
&=&\frac{1}{2}%
\begin{pmatrix}
1+v_{z}(t) & v_{x}(t)-iv_{y}(t) \\ 
v_{x}(t)+iv_{y}(t) & 1-v_{z}(t)%
\end{pmatrix}%
.  \label{I+vs}
\end{eqnarray}%
We assume that the initial state is a product state, i.e., 
\begin{equation}
\rho (0)=\rho _{S}(0)\otimes \rho _{B},
\end{equation}%
and that the bath is initially in the Gibbs thermal state at a temperature $T
$ 
\begin{equation}
\rho _{B}=\exp (-H_{B}/kT)/\mathrm{Tr}[\exp (-H_{B}/kT)],  \label{eq:rhoB0}
\end{equation}%
where $k$ is the Boltzmann constant. Since $\rho _{B}$ commutes with the
interaction Hamiltonian $H_{I}$, the bath state is stationary throughout the
dynamics:\ $\rho _{B}(t)=\rho _{B}$. Finally, the bath spectral density
function is defined as usual as%
\begin{equation}
J(\Omega )=\sum_{n}|g_{n}|^{2}\delta (\Omega -\Omega _{n}).  \label{eq:J}
\end{equation}

\subsection{Exact solution for the system-spin dynamics}

We first shift the system Hamiltonian in the following way: 
\begin{eqnarray}
H_{S} &\mapsto &H_{S}+\theta I,  \notag \\
\theta &\equiv &\mathrm{Tr}\{\sum_{n}g_{n}\sigma _{n}^{z}\rho _{B}\}.
\label{theta}
\end{eqnarray}%
As a consequence the interaction Hamiltonian is modified from Eq. (\ref%
{eq:HI}) to 
\begin{equation}
H_{I}^{\prime }\mapsto H_{I}=\alpha \sigma ^{z}\otimes B,
\end{equation}%
where 
\begin{equation}
B\equiv \sum_{n}g_{n}\sigma _{n}^{z}-\theta I_{B}.  \label{Bcomp}
\end{equation}%
This shift is performed because now $\mathrm{Tr}_{B}[H_{I},\rho (0)]=0$, or
equivalently 
\begin{equation}
\mathrm{Tr}_{B}\{B\rho _{B}\}=0.
\end{equation}%
This property will simplify our calculations later when we consider
approximation techniques in Sec. III. Now, we derive the exact solution for
the reduced density operator $\rho _{S}$ corresponding to the system. We do
this in two different ways. The Kraus operator sum representation is a
standard description of the dynamics of a system initially decoupled from
its environment and it will also be helpful in studying the coarse-graining
approach to the quantum semigroup master equation. The second method is
computationally more effective and is helpful in obtaining analytical
expressions for $N\gg 1$.

\subsubsection{Exact Solution in the Kraus Representation}

In the Kraus representation the system state at any given time can be
written as 
\begin{equation}
\rho _{S}(t)=\sum_{i,j}K_{ij}\rho _{S}(0)K_{ij}^{\dag },  \label{KrausForm}
\end{equation}%
where the Kraus operators satisfy $\sum_{ij}K_{ij}^{\dag }K_{ij}=I_{S}$ \cite%
{Kraus:83}. These operators can be expressed easily in the eigenbasis of the
initial state of the bath density operator as 
\begin{equation}
K_{ij}=\sqrt{\lambda _{i}}\langle j|\exp (-iH_{I}t)|i\rangle ,
\end{equation}%
where the bath density operator at the initial time is $\rho
_{B}(0)=\sum_{i}\lambda _{i}|i\rangle \langle i|$. For the Gibbs thermal
state chosen here, the eigenbasis is the $N$-fold tensor product of the $%
\sigma ^{z}$ basis. In this basis 
\begin{equation}
\rho _{B}=\sum_{l}\frac{\exp (-\beta E_{l})}{Z}|l\rangle \langle l|,
\end{equation}%
where $\beta =1/kT$. Here 
\begin{equation}
E_{l}=\sum_{n=1}^{N}\frac{1}{2}\hbar \Omega _{n}(-1)^{l_{n}},  \label{eq:El}
\end{equation}%
is the energy of each eigenstate $|l\rangle $, where $l=l_{1}l_{2}\dots
l_{n} $ is the binary expansion of the integer $l$, and the partition
function is $Z=\sum_{l}\exp (-\beta E_{l})$. Therefore, the Kraus operators
become 
\begin{equation}
K_{ij}=\sqrt{\lambda _{i}}\exp (-it\alpha \tilde{E}_{i}\sigma ^{z})\delta
_{ij},
\end{equation}%
where 
\begin{equation}
\tilde{E}_{i}=\langle i|B|i\rangle =\sum_{n=1}^{N}g_{n}(-1)^{i_{n}}-\mathrm{%
Tr}\{\sum_{n}g_{n}\sigma _{n}^{z}\rho _{B}\},  \label{eq:Eitilde}
\end{equation}%
and $\lambda _{i}=\exp (-\beta E_{i})/Z$. Substituting this expression for $%
K_{ij}$ into Eq. (\ref{KrausForm}) and writing the system state in the Bloch
vector form given in Eq. (\ref{I+vs}), we obtain 
\begin{eqnarray}
v_{x}(t) &=&v_{x}(0)C(t)-v_{y}(0)S(t),  \notag \\
v_{y}(t) &=&v_{x}(0)S(t)+v_{y}(0)C(t),  \label{ExactSoln} \\
v_{z}(t) &=&v_{z}(0),  \notag
\end{eqnarray}%
where%
\begin{eqnarray}
C(t) &=&\sum_{i}\lambda _{i}\cos 2\alpha \tilde{E}_{i}t,  \notag \\
S(t) &=&\sum_{i}\lambda _{i}\sin 2\alpha \tilde{E}_{i}t.  \label{CSeqn}
\end{eqnarray}

The equations (\ref{ExactSoln}) are the exact solution to the system
dynamics of the above spin bath model. We see that the evolution of the
Bloch vector is a linear combination of rotations around the $z$ axis. This
evolution reflects the symmetry of the interaction Hamiltonian which is
diagonal in the $z$ basis. By inverting Eqs. (\ref{ExactSoln}) for $v_{x}(0)$
or $v_{y}(0)$, we see that the Kraus map is irreversible when $%
C(t)^{2}+S(t)^{2}=0$. This will become important below, when we discuss the
validity of the time-convolutionless approximation.

\subsubsection{Alternative Exact Solution}

Another way to derive the exact solution which is computationally more
useful is the following. Since all $\sigma _{n}^{z}$ commute, the initial
bath density matrix factors and can be written as 
\begin{eqnarray}
\rho _{B} &=&\bigotimes\limits_{n=1}^{N}\frac{\exp \left( -\frac{\Omega _{n}%
}{2kT}\sigma _{n}^{z}\right) }{\mathrm{Tr}\left[ \exp \left( -\frac{\Omega
_{n}}{2kT}\sigma _{n}^{z}\right) \right] }  \notag \\
&=&\bigotimes\limits_{n=1}^{N}\frac{1}{2}\left( I+\beta _{n}\sigma
_{n}^{z}\right) \equiv \prod_{n=1}^{N}\rho _{n},  \label{eq_rho_B_inter}
\end{eqnarray}%
where 
\begin{equation}
\beta _{n}=\tanh \left( -\frac{\Omega _{n}}{2kT}\right) ,
\end{equation}%
and $-1\leq \beta _{n}\leq 1$. Using this, we obtain an expression for $%
\theta $ defined in Eq. (\ref{theta}) 
\begin{eqnarray}
\theta  &=&\mathrm{Tr}\{\sum_{n=1}^{N}g_{n}\sigma
_{n}^{z}\bigotimes\limits_{m=1}^{N}\frac{1}{2}(I+\beta _{m}\sigma _{m}^{z})\}
\notag \\
&=&\sum_{n=1}^{N}g_{n}\mathrm{Tr}\{\frac{1}{2}(\sigma _{n}^{z}+\beta
_{n}I)\}\prod\limits_{m\neq n}\mathrm{Tr}\{\frac{1}{2}(I+\beta _{m}\sigma
_{m}^{z})\}  \notag \\
&=&\sum_{n=1}^{N}g_{n}\beta _{n}.  \label{eq:theta}
\end{eqnarray}%
The evolution of the system density matrix in the interaction picture is 
\begin{equation}
\rho _{S}(t)=\mathrm{Tr}_{B}\{e^{-iH_{I}t}\rho (0)e^{iH_{I}t}\}.
\end{equation}%
In terms of the system density matrix elements in the computational basis $%
\{|0\rangle ,|1\rangle \}$ (which is an eigenbasis of $\sigma ^{z}$ in $%
H_{I}=\alpha \sigma ^{z}\otimes B$), we have 
\begin{eqnarray}
\langle j|\rho _{S}(t)|k\rangle  &=&\langle j|\mathrm{Tr}_{B}\{e^{-iH_{I}t} 
\notag \\
&\times &\rho _{S}(0)\bigotimes\limits_{m=1}^{N}\rho
_{m}e^{iH_{I}t}\}|k\rangle   \notag \\
&=&\mathrm{Tr}_{B}\{e^{-i\alpha \langle j|\sigma ^{z}|j\rangle Bt}  \notag \\
&\times &\langle j|\rho _{S}(0)|k\rangle \bigotimes\limits_{m=1}^{N}\rho
_{m}e^{+i\alpha \langle k|\sigma ^{z}|k\rangle Bt}\}.  \notag
\end{eqnarray}%
Let us substitute $\langle j|\sigma ^{z}|j\rangle =(-1)^{j}$ and rewrite 
\begin{eqnarray}
e^{-i\alpha \langle j|\sigma ^{z}|j\rangle Bt} &=&e^{-i\alpha (-1)^{j}\left(
\sum_{l=1}^{N}g_{l}\sigma _{l}^{z}-\theta I\right) t}  \notag \\
&=&\bigotimes\limits_{l=1}^{N}e^{-i(-1)^{j}\alpha \left( g_{l}\sigma
_{l}^{z}-\frac{\theta }{N}I\right) t}.  \notag
\end{eqnarray}%
Since all the matrices are diagonal, they commute and we can collect the
terms by qubits: 
\begin{eqnarray}
\langle j|\rho _{S}(t)|k\rangle  &=&\langle j|\rho _{S}(0)|k\rangle   \notag
\\
&\times &\mathrm{Tr}\{\bigotimes\limits_{m=1}^{N}e^{-i\left[
(-1)^{j}-(-1)^{k}\right] \alpha \left( g_{l}\sigma _{l}^{z}-\frac{\theta }{N}%
I\right) t}\rho _{n}\}.  \notag
\end{eqnarray}%
Let us denote $(-1)^{j}-(-1)^{k}=2\epsilon _{jk}$. The trace can be easily
computed to be 
\begin{eqnarray*}
&\prod_{n=1}^{N}&\mathrm{Tr}\{e^{-i2\epsilon _{jk}\alpha \left( g_{n}\sigma
_{n}^{z}-\frac{\theta }{N}I\right) t}\tfrac{1}{2}(I+\beta _{n}\sigma
_{n}^{z})\} \\
&=&\prod_{n=1}^{N}e^{i2\epsilon _{jk}\alpha \frac{\theta }{N}t}\left[ \cos
(2\epsilon _{jk}\alpha g_{n}t)-i\beta _{n}\sin (2\epsilon _{jk}\alpha g_{n}t)%
\right] .
\end{eqnarray*}%
Thus the final expression for the system density matrix elements is 
\begin{eqnarray}
\langle j|\rho _{S}(t)|k\rangle  &=&\langle j|\rho _{S}(0)|k\rangle
e^{i2\epsilon _{jk}\alpha \theta t}  \notag \\
&\times &\prod_{n=1}^{N}\left[ \cos (2\epsilon _{jk}\alpha g_{n}t)-i\beta
_{n}\sin (2\epsilon _{jk}\alpha g_{n}t)\right] .  \notag
\end{eqnarray}%
Notice that $\epsilon _{00}=\epsilon _{11}=0$, hence the diagonal matrix
elements do not depend on time as before: 
\begin{gather*}
\langle 0|\rho _{S}(t)|0\rangle =\langle 0|\rho _{S}(0)|0\rangle , \\
\langle 1|\rho _{S}(t)|1\rangle =\langle 1|\rho _{S}(0)|1\rangle .
\end{gather*}%
For the off-diagonal matrix elements $\epsilon _{01}=1$, $\epsilon _{10}=-1$%
, and the evolution is described by 
\begin{eqnarray}
\langle 0|\rho _{S}(t)|1\rangle  &=&\langle 0|\rho _{S}(0)|1\rangle f(t), 
\notag \\
\langle 1|\rho _{S}(t)|0\rangle  &=&\langle 1|\rho _{S}(0)|0\rangle f^{\ast
}(t),  \label{ExactSoln2}
\end{eqnarray}%
where 
\begin{equation}
f(t)=e^{i2\alpha \theta t}\prod_{n=1}^{N}\left[ \cos (2\alpha g_{n}t)-i\beta
_{n}\sin (2\alpha g_{n}t)\right] .  \label{prod}
\end{equation}%
In terms of the Bloch vector components, this can be written in the form of
Eq. (\ref{ExactSoln}), where 
\begin{eqnarray}
C(t) &=&(f(t)+f^{\ast }(t))/2,  \notag \\
S(t) &=&(f(t)-f^{\ast }(t))/2i.  \label{eq:CS}
\end{eqnarray}

\subsection{Limiting cases}

\subsubsection{Short Times}

Consider the evolution for short times where $\alpha t\ll 1$. Then 
\begin{eqnarray}
\lefteqn{\left\vert \prod_{n=1}^{N}\left[ \cos (2\alpha g_{n}t)\pm i\beta
_{n}\sin (2\alpha g_{n}t)\right] \right\vert }  \notag \\
&=&\prod_{n=1}^{N}\sqrt{1-(1-\beta _{n}^{2})\sin ^{2}(2\alpha g_{n}t)} 
\notag \\
&\approx &\prod_{n=1}^{N}[1-2(1-\beta _{n}^{2})(\alpha g_{n}t)^{2}]  \notag
\\
&\approx &1-2\left[ \alpha ^{2}\sum_{n=1}^{N}g_{n}^{2}(1-\beta _{n}^{2})%
\right] t^{2}  \notag \\
&\approx &\exp [-2(\alpha t)^{2}Q_{2}],  \label{eq:f-approx}
\end{eqnarray}%
where (see Appendix \ref{app:A}) 
\begin{eqnarray}
Q_{2} &\equiv &\mathrm{Tr}\{B^{2}\rho _{B}\}=\sum_{n=1}^{N}g_{n}^{2}(1-\beta
_{n}^{2})  \notag \\
&=&\int_{-\infty }^{\infty }\frac{2J(\Omega )}{1+\cosh (\frac{\Omega }{kT})}%
\mathrm{d}\Omega .  \label{Q_2}
\end{eqnarray}%
Note that for the above approximation to be valid, we need $2(\alpha
t)^{2}Q_{2}\ll 1$. The total phase of $f(t)$ in Eq. (\ref{prod}) is 
\begin{eqnarray}
\phi  &\approx &2\theta \alpha t+\sum_{n=1}^{N}(-\beta _{n}2\alpha g_{n}t) 
\notag \\
&=&2\theta \alpha t-2\alpha \left( \sum_{n=1}^{N}g_{n}\beta _{n}\right) t=0,
\end{eqnarray}%
where we have used Eq. (\ref{eq:theta}). Thus, the off-diagonal elements of
the system density matrix become 
\begin{eqnarray}
\rho _{S}^{01}(t) &\approx &\rho _{S}^{01}(0)e^{-2(\alpha t)^{2}Q_{2}}, 
\notag \\
\rho _{S}^{10}(t) &\approx &\rho _{S}^{10}(0)e^{-2(\alpha t)^{2}Q_{2}}.
\end{eqnarray}

Finally, the dynamics of the Bloch vector components are: 
\begin{eqnarray}
v_{x,y}(t) &\approx &v_{x,y}(0)e^{-2(\alpha t)^{2}Q_{2}},  \notag \\
v_{z}(t) &=&v_{z}(t).  \label{shorttimes}
\end{eqnarray}%
This represents the well known behavior \cite{NNP96} of the evolution of an
open quantum system in the Zeno regime. In this regime coherence does not
decay exponentially but is initially flat, as is the case here due to the
vanishing time derivative of $\rho _{S}^{01}(t)$ at $t=0$. As we will see in
Sec. III, the dynamics in the Born approximation (which is also the second
order time-convolutionless approximation) exactly matches the last result.

\subsubsection{Large $N$}

When $N\gg 1$ and the values of $g_{n}$ are random, then the different terms
in the product of Eq. (\ref{prod}) are smaller than $1$ most of the time and
have recurrences at different times. Therefore, we expect the function $f(t)$
to be close to zero in magnitude for most of the time and full recurrences,
if they exist, to be extremely rare. When $g_{n}$ are equal and so are $%
\Omega _{n}$, then partial recurrences occur periodically, independently of $%
N$. Full recurrences occur with a period which grows at least as fast as $N$%
. This can be argued from Eq. (\ref{ExactSoln}) by imposing the condition
that the arguments of all the cosines and sines are simultaneously equal to
an integer multiple of $2\pi $. When $J(\Omega )$ has a narrow high peak,
e.g., one $g_{n}$ is much larger than the others, then the corresponding
terms in the products in Eq. (\ref{prod}) oscillate faster than the rate at
which the whole product decays. This is effectively a modulation of the
decay.

\subsubsection{Discontinuous spectral density co-domain}

As can be seen from Eq.~(\ref{prod}), the coupling constants $g_{n}$
determine the oscillation periods of the product terms, while the
temperature factors $\beta _{n}$ determine their modulation depths. If the
codomain of spectral density is not continuous, i.e. it can be split into
nonoverlapping intervals $G_{j}$, $j=1,...,J$, then Eq.~(\ref{prod}) can be
represented in the following form: 
\begin{equation}
f(t)=e^{i2\alpha \theta t}P_{1}(t)P_{2}(t)\dots P_{J}(t),
\end{equation}%
where 
\begin{equation}
P_{j}(t)=\prod_{g_{n}\in G_{j}}\big[\cos (2\alpha g_{n}t)-i\beta _{n}\sin
(2\alpha g_{n}t)\big].
\end{equation}%
In this case, if $G_{j}$ are separated by large enough gaps, the evolution
rates of different $P_{j}(t)$ can be significantly different. This is
particularly noticeable if one $P_{j}(t)$ undergoes partial recurrences
while another $P_{j^{\prime }}(t)$ slowly decays.

For example, one can envision a situation with two intervals such that one
term shows frequent partial recurrences that slowly decay with time, while
the other term decays faster, but at times larger than the recurrence time.
The overall evolution then consists in a small number of fast partial
recurrences. In an extreme case, when one $g_{n}$ is much larger then the
others, this results in an infinite harmonic modulation of the decay with
depth dependent on $\beta _{n}$, i.e., on temperature.

\subsubsection{Alternating signs}

If the bath has the property that every bath qubit $m$ has a pair $-m$ with
the same frequency $\Omega _{-m}=\Omega _{m}$, but opposite coupling
constant $g_{-m}=-g_{m}$, the exact solution can be simplified. First, $\beta
_{-m}=\beta _{m}$, and $\theta =0$. Next, Eq.~(\ref{prod}) becomes 
\begin{eqnarray}
f(t) &=&\prod_{m=1}^{N/2}\big[\cos (2\alpha g_{m}t)-i\beta _{m}\sin (2\alpha
g_{m}t)\big]\times   \notag \\
&&\big[\cos (2\alpha g_{-m}t)-i\beta _{-m}\sin (2\alpha g_{-m}t)\big]  \notag
\\
&=&\prod_{m=1}^{N/2}\big[\cos ^{2}(2\alpha g_{m}t)+\beta _{m}^{2}\sin
^{2}(2\alpha g_{m}t)\big].
\end{eqnarray}%
This function is real, thus Eq.~(\ref{eq:CS}) becomes $C(t)=f(t),S(t)=0$, so
that $v_{x}(t)=v_{x}(0)f(t)$ and $v_{y}(t)=v_{y}(0)f(t)$. The exact solution
is then symmetric under the interchange $v_{x}\leftrightarrow v_{y}$, a
property shared by all the second order approximate solutions considered
below, as well as the post-Markovian master equation. The limiting case Eq.~(%
\ref{eq:f-approx}) remains unchanged, and since $Q_{2}$ depends on $g_{n}^{2}
$, but not $g_{n}$, it and all second order approximations also remain
unchanged. In the special case $|g_{m}|=g$, the exact solution exhibits full
recurrences with period $T=\pi /\alpha g$.

\section{Approximation methods}

In this section we discuss the performance of different approximation
methods developed in the open quantum systems literature \cite%
{Alicki:87,Breuer:book}. The corresponding master equations for the system
density matrix can be derived explicitly and since the model considered here
is exactly solvable, we can compare the appoximations to the exact dynamics.
We use the Bloch vector representation and since the $z$ component has no
dynamics, a fact which is reflected in all the master equations, we omit it
from our comparisons.

\subsection{Born and Born-Markov approximations}

Both the Born and Born-Markov approximations are second order in the
coupling strength $\alpha $.

\subsubsection{Born approximation}

The Born approximation is equivalent to a truncation of the Nakajima-Zwanzig
projection operator method at the second order, which is discussed in detail
in Sec. III B. The Born approximation is given by the following
integro-differential master equation:%
\begin{equation}
\dot{\rho}_{S}(t)=-\int_{0}^{t}\mathrm{Tr}_{B}\{[H_{I}(t),[H_{I}(s),\rho
_{S}(s)\otimes \rho _{B}]]\}\text{d}s.
\end{equation}%
Since in our case the interaction Hamiltonian is time-independent, the
integral becomes easy to solve. We obtain 
\begin{equation}
\dot{\rho}_{S}(t)=-2\alpha ^{2}Q_{2}\int_{0}^{t}(\rho _{S}(s)-\sigma
^{z}\rho _{S}(s)\sigma ^{z})\text{d}s,
\end{equation}%
where $Q_{2}$ is the second order bath correlation function in Eq. (\ref{Q_2}%
). Writing $\rho _{S}(t)$ in terms of Bloch vectors as $(I+\vec{v}\cdot \vec{%
\sigma})/2$ [Eq. (\ref{I+vs})], we obtain the following integro-differential
equations: 
\begin{eqnarray}
\dot{v}_{x,y}(t) &=&-4\alpha ^{2}Q_{2}\int_{0}^{t}v_{x,y}(s)\text{d}s.
\label{BornApprox}
\end{eqnarray}

These equations can be solved by taking the Laplace transform of the
variables. The equations become 
\begin{equation}
sV_{x,y}(s)-v_{x,y}(0)=-4\alpha ^{2}Q_{2}\frac{V_{x,y}(s)}{s},
\end{equation}%
where $V_{x,y}(s)$ is the Laplace transform of $v_{x,y}(t)$. This gives 
\begin{equation}
V_{x,y}(s)=\frac{v_{x,y}(0)s}{s^{2}+4Q_{2}\alpha ^{2}},
\end{equation}%
which can be readily solved by taking the inverse Laplace transform. Doing
so, we obtain the solution of the Born master equation for our model: 
\begin{eqnarray}
v_{x,y}(t) &=&v_{x,y}(0)\cos (2\alpha \sqrt{Q_{2}}t).  \label{Born}
\end{eqnarray}%
Note that this solution is symmetric under the interchange $%
v_{x}\leftrightarrow v_{y}$, but the exact dynamics in Eq. (\ref{ExactSoln})
does not have this symmetry. The exact dynamics respects the symmetry: $%
v_{x}\rightarrow v_{y}$ and $v_{y}\rightarrow -v_{x}$, which is a symmetry
of the Hamiltonian. This means that higher order corrections are required to
break the symmetry $v_{x}\leftrightarrow v_{y}$ in order to approximate the
exact solution more closely.

One often makes the substitution $v_{x,y}(t)$ for $v_{x,y}(s)$ in Eq. (\ref%
{BornApprox}) since the integro-differential equation obtained in other
models may not be as easily solvable. This approximation, which is valid for
short times, yields 
\begin{eqnarray}
\dot{v}_{x,y}(t) &=&-4\alpha ^{2}Q_{2}tv_{x,y}(t),
\end{eqnarray}%
which gives 
\begin{eqnarray}
v_{x,y}(t) &=&v_{x,y}(0)\exp (-2Q_{2}\alpha ^{2}t^{2}),  \label{TCL2}
\end{eqnarray}%
i.e., we recover Eq. (\ref{shorttimes}). This is the same solution obtained
in the second order approximation using the time-convolutionless (TCL)
projection method discussed in Sec. III B.

\subsubsection{Born-Markov approximation}

In order to obtain the Born-Markov approximation, we use the following
quantities \cite{Breuer:book}[Ch.3]: 
\begin{eqnarray}
R(\omega ) &=&\sum_{E_{2}-E_{1}=\omega }P_{E_{1}}\sigma ^{z}P_{E_{2}}, 
\notag \\
\Gamma (\omega ) &=&\alpha ^{2}\int_{0}^{\infty }e^{i\omega s}Q_{2}\text{d}s,
\notag \\
H_{L} &=&\sum_{\omega }T(\omega )R(\omega )^{\dag }R(\omega ),
\end{eqnarray}%
where $T(\omega )=(\Gamma (\omega )-\Gamma (\omega )^{\ast })/2i$, $E_{i}$
is an eigenvalue of the system Hamiltonian $H_{S}$, and $P_{E_{i}}$ is the
projector onto the eigenspace corresponding to this eigenvalue. In our case $%
H_{S}$ is diagonal in the eigenbasis of $\sigma ^{z}$, and only $\omega =0$
is relevant. This leads to $R(0)=\sigma ^{z}$ and $\Gamma (0)=\alpha
^{2}\int_{0}^{\infty }Q_{2}\text{d}t$. Since $\Gamma (0)$ is real, we have $%
T(0)=0.$ Hence the Lamb shift Hamiltonian $H_{L}=0$, and the Lindblad form
of the Born-Markov approximation is 
\begin{equation}
\dot{\rho}_{S}(t)=\gamma (\sigma ^{z}\rho _{S}\sigma ^{z}-\rho _{S}),
\label{Lindblad}
\end{equation}%
where $\gamma =\Gamma (0)+\Gamma (0)^{\ast }=2\alpha ^{2}\int_{0}^{\infty
}Q_{2}$d$t$. But note that $Q_{2}=\mathrm{Tr}_{B}\{B^{2}\rho _{B}\}$ does
not depend on time. This means that $\Gamma $ and hence $\gamma $ are both
infinite. Thus the Born-Markov approximation is not valid for this model and
the main reason for this is the time independence of the bath correlation
functions. The dynamics is inherently non-Markovian.

A different approach to the derivation of a Markovian semigroup master
equation was proposed in \cite{Lidar:CP01}. In this approach, a Lindblad
equation is derived from the Kraus operator-sum representation by a
coarse-graining procedure defined in terms of a phenomenological
coarse-graining time scale $\tau $. The general form of the equation is: 
\begin{eqnarray}
\frac{\partial \rho (t)}{\partial t} &=&-i[\langle \dot{Q}\rangle _{\tau
},\rho (t)]  \notag \\
&+&\frac{1}{2}\sum_{\alpha ,\beta =1}^{M}\langle \dot{\chi}_{\alpha ,\beta
}\rangle _{\tau }([A_{\alpha },\rho (t)A_{\beta }^{\dagger }]+[A_{\alpha
}\rho (t),A_{\beta }^{\dagger }]),  \notag
\end{eqnarray}%
where the operators $A_{0}=I$ and $A_{\alpha },\alpha =1,...,M$ form an
arbitrary fixed operator basis in which the Kraus operators (\ref{KrausForm}%
) can be expanded as 
\begin{equation}
K_{i}=\sum_{\alpha =0}^{M}b_{i\alpha }A_{\alpha }.
\end{equation}%
The quantities $\chi _{\alpha ,\beta }(t)$ and $Q(t)$ are defined through 
\begin{equation}
\chi _{\alpha ,\beta }(t)=\sum_{i}b_{i\alpha }(t)b_{i\beta }^{\ast }(t),
\end{equation}%
\begin{equation}
Q(t)=\frac{i}{2}\sum_{\alpha =1}^{M}(\chi _{\alpha 0}(t)K_{\alpha }-\chi
_{0\alpha }(t)K_{\alpha }^{\dagger }),
\end{equation}%
and 
\begin{equation}
\langle X\rangle _{\tau }=\frac{1}{\tau }\int_{0}^{\tau }X(s)ds.
\end{equation}%
For our problem we find 
\begin{equation}
\frac{\partial \rho (t)}{\partial t}=-i\tilde{\omega}[\sigma _{Z},\rho (t)]+%
\tilde{\gamma}(\sigma _{Z}\rho (t)\sigma _{Z}-\rho (t)),  \label{coarsegrain}
\end{equation}%
where 
\begin{equation}
\tilde{\omega}=\frac{1}{2\tau }S(\tau )
\end{equation}%
and 
\begin{equation}
\tilde{\gamma}=\frac{1}{2\tau }(1-C(\tau ))
\end{equation}%
with $C(t)$ and $S(t)$ defined in Eq. (\ref{CSeqn}). In order for this
approximation to be justified, it is required that the coarse-graining time
scale $\tau $ be much larger than any characteristic time scale of the bath 
\cite{Lidar:CP01}. However, in our case the bath correlation time is
infinite which, once again, shows the inapplicability of the Markovian
approximation. This is further supported by the performance of the optimal
solution that one can achieve by varying $\tau $, which is discussed in Sec.
IV. There we numerically examine the average trace-distance between the
solution to Eq. (\ref{coarsegrain}) and the exact solution as a function of $%
\tau $. The average is taken over a time $T$, which is greater than the
decay time of the exact solution. We determine an optimal $\tau $ for which
the average trace distance is minimum and then determine the approximate
solution. The solution of Eq. (\ref{coarsegrain}) for a particular $\tau $
in terms of the Bloch vector components is 
\begin{eqnarray}
v_{x}(t) &=&v_{x}(0)\tilde{C}_{\tau }(t)+v_{y}(0)\tilde{S}_{\tau }(t)  \notag
\\
v_{y}(t) &=&v_{y}(0)\tilde{C}_{\tau }(t)-v_{x}(0)\tilde{S}_{\tau }(t),
\end{eqnarray}%
where $\tilde{C}_{\tau }(t)=e^{-\tilde{\gamma}(\tau )t}\cos (\tilde{\omega}%
(\tau )t)$ and $\tilde{S}_{\tau }(t)=e^{-\tilde{\gamma}(\tau )t}\sin (\tilde{%
\omega}(\tau )t)$. The average trace distance as a function of $\tau $ is
given by, 
\begin{eqnarray}
&\bar{D}&(\rho _{\mathrm{exact}},\rho _{\mathrm{CG}})\equiv \frac{1}{2}%
\mathrm{Tr}|\rho _{\mathrm{exact}}-\rho _{\mathrm{CG}}|  \notag \\
&=&\frac{1}{2T}\sum_{t=0}^{T}\sqrt{(C(t)-\tilde{C}(t))^{2}+(S(t)-\tilde{S}%
(t))^{2}}  \notag \\
&\times &\sqrt{v_{x}(0)^{2}+v_{y}(0)^{2}},
\end{eqnarray}%
where $\rho _{\mathrm{CG}}$ represents the coarse-grained solution and where 
$|X|=\sqrt{X^{\dag }X}$. The results are presented in Sec. IV. Next we
consider the Nakajima-Zwanzig (NZ) and the time-convolutionless (TCL) master
equations for higher order approximations.

\subsection{NZ and TCL master equations}

Using projection operators one can obtain approximate non-Markovian master
equations to higher orders in $\alpha t$. A projection is defined as
follows, 
\begin{equation}
\mathcal{P}\rho=\mathrm{Tr}_B\{\rho\}\otimes\rho_B ,
\end{equation}
and serves to focus on the ``relevant dynamics'' (of the system) by removing
the bath (a recent generalization is discussed in Ref. \cite{Breuer:07}).
The choice of $\rho_B$ is somewhat arbitrary and can be taken to be $%
\rho_B(0)$ which significantly simplifies the calculations. Using the
notation introduced in \cite{BBP04}, define 
\begin{equation}
\left\la \mathcal{S}\right\ra\equiv \mathcal{P}\mathcal{S}\mathcal{P}
\end{equation}
for any superoperator $\mathcal{S}$. Thus $\left\la\mathcal{S}^n\right\ra$
denote the moments of the superoperator. Note that for the Liouvillian
superoperator, $\left\la\mathcal{L}\right\ra=0$ by virtue of the fact that $%
\mathrm{Tr}_B\{B\rho_B(0)\}=0$ (see \cite{Breuer:book}). Since we assume
that the initial state is a product state, both the NZ and TCL equations are
homogeneous equations. The NZ master equation is an integro-differential
equation with a memory kernel $\mathcal{N}(t,s)$ and is given by 
\begin{equation}
\dot{\rho}_S(t)\otimes\rho_B=\int_0^t\mathcal{N}(t,s)\rho_S(s)\otimes\rho_B 
\text{d}s .
\end{equation}
The TCL master equation is a time-local equation given by 
\begin{equation}
\dot{\rho}_S(t)\otimes\rho_B=\mathcal{K}(t)\rho_S(t)\otimes\rho_B .
\end{equation}
When these equations are expanded in $\alpha t$ and solved we obtain the
higher order corrections. When the interaction Hamiltonian is time
independent (as in our case), the above equations simplify to 
\begin{equation}  \label{NZ}
\int_0^t\mathcal{N}(t,s)\rho_S(s)\otimes\rho_B \text{d}s=\sum_{n=1}^\infty
\alpha^n \mathcal{I}_n(t,s)\left\la\mathcal{L}^n \right\ra_{pc}\rho_S(s)
\end{equation}
and 
\begin{equation}  \label{TCL}
\mathcal{K}(t)=\sum_{n=1}^\infty \alpha^n \frac{t^{n-1}}{(n-1)!}\left\la 
\mathcal{L}^n\right\ra_{oc}
\end{equation}
for the NZ and TCL equations, respectively, where the time-ordered integral
operator $\mathcal{I}_n(t,s)$ is defined as 
\begin{equation}
\mathcal{I}_n(t,s)\equiv \int_0^t\text{d}t_1\int_0^{t_1}\text{d}%
t_2\cdots\int_0^{t_{n-2}}\text{d}s .
\end{equation}
The definitions of the partial cumulants $\left\la\mathcal{L}\right\ra_{pc}$
and the ordered cumulants $\left\la\mathcal{L}\right\ra_{oc}$ are given in
Refs. \cite{ShiAri80,Royer:72,Kam74}. For our model we have 
\begin{equation}
\left\la\mathcal{L}\right\ra_{pc}=\left\la\mathcal{L}\right\ra_{oc}=0 ,
\end{equation}
and 
\begin{eqnarray}
\left\la\mathcal{L}^2\right\ra_{pc}=\left\la\mathcal{L}^2\right\ra  \notag \\
\left\la\mathcal{L}^2\right\ra_{oc}=\left\la\mathcal{L}^2\right\ra  \notag \\
\left\la\mathcal{L}^3\right\ra_{pc}=\left\la\mathcal{L}^3\right\ra  \notag \\
\left\la\mathcal{L}^3\right\ra_{oc}=\left\la\mathcal{L}^3\right\ra  \notag \\
\left\la\mathcal{L}^4\right\ra_{pc}=\left\la\mathcal{L}^4\right\ra - \left\la%
\mathcal{L}^2\right\ra^2  \notag \\
\left\la\mathcal{L}^4\right\ra_{oc}=\left\la\mathcal{L}^4\right\ra -
3\left\la\mathcal{L}^2\right\ra^2 .  \label{eq:cumulants}
\end{eqnarray}
Explicit expressions for these quantities are given in Appendix \ref{app:B}.
Substituting these into the NZ and TCL equations (\ref{NZ}) and (\ref{TCL}),
we obtain what we refer to below as the NZ$n$ and TCL$n$ master equations,
with $n=2,3,4$. These approximate master equations are, respectively,
second, third and fourth order in the coupling constant $\alpha$, and they
can be solved analytically. The second order solution of the NZ equation
(NZ2) is exactly the Born approximation and the solution is given in Eq. (%
\ref{Born}). The third order NZ master equation is given by 
\begin{eqnarray}
\dot{\rho}_S(t)&=&-2\alpha^2Q_2 \mathcal{I}_2(t,s)
(\rho_S(s)-\sigma^z\rho_S(s)\sigma^z)  \notag \\
&+& i4\alpha^3Q_3\mathcal{I}_3(t,s)(\sigma^z\rho_S(s)-\rho_S(s)\sigma^z) ,
\end{eqnarray}
and the fourth order is 
\begin{eqnarray}
\dot{\rho}_S(t)&=&-2\alpha^2Q_2 \mathcal{I}_2(t,s)
(\rho_S(s)-\sigma^z\rho_S(s)\sigma^z)  \notag \\
&+& i4\alpha^3Q_3\mathcal{I}_3(t,s)(\sigma^z\rho_S(s)-\rho_S(s)\sigma^z) 
\notag \\
&+& 8\alpha^4(Q_4-Q_2^2)\mathcal{I}_4(t,s)(\rho_S(s)-\sigma^z\rho_S(s)%
\sigma^z).  \notag \\
\end{eqnarray}
These equations are equivalent to, respectively, 6th and 8th order
differential equations (with constant coefficients) and are difficult to
solve analytically. The results we present in the next section were
therefore obtained numerically.

The situation is simpler in the TCL approach. The second order TCL equation
is given by 
\begin{eqnarray}
\dot{\rho}_S(t)&=&-\alpha^2t\mathrm{Tr}_B\{[H_I,[H_I,\rho_S(t)\otimes%
\rho_B(0)]]\}  \notag \\
&=& -2\alpha^2tQ_2(\rho_S(t)-\sigma^z\rho_S(t)\sigma^z) ,
\end{eqnarray}
whose solution is as given in Eq. (\ref{TCL2}) in terms of Bloch vector
components. For TCL3 we find 
\begin{eqnarray}
\dot{\rho}_S(t)&=&-2\alpha^2 tQ_2(\rho_S(t)-\sigma_z\rho_S(t)\sigma_z) 
\notag \\
&+& 4iQ_3\alpha^3\frac{t^2}{2}(\sigma_z\rho_S(t)-\rho_S(t)\sigma_z),
\end{eqnarray}
and for TCL4 we find 
\begin{eqnarray}
\dot{\rho}_S(t) &=& [-2\alpha^2 tQ_2 + (8Q_4-24Q_2^2)\alpha^4\frac{t^3}{6}] 
\notag \\
&\times&(\rho_S(t)-\sigma_z\rho_S(t)\sigma_z)  \notag \\
&+& 4iQ_3\alpha^3\frac{t^2}{2}(\sigma_z\rho_S(t)-\rho_S(t)\sigma_z) .
\end{eqnarray}
These equation can be solved analytically, and the solutions to the third
and fourth order TCL equations are given by 
\begin{eqnarray}  \label{TCL3}
v_x(t)&=&f_n(\alpha t)\left [v_x(0)\cos(g(t))+v_y(0)\sin(g(t))\right ], 
\notag \\
v_y(t)&=&f_n(\alpha t)\left [v_y(0)\cos(g(t))-v_x(0)\sin(g(t))\right ] . 
\notag \\
\end{eqnarray}
where $g(t)=4Q_3\alpha^3t^3/3$, $f_3(\alpha t)=\exp(-2Q_2\alpha^2t^2)$
(TCL3) and $f_4(\alpha t)=\exp(-2Q_2\alpha^2t^2+(2Q_4-6Q_2^2)\alpha^4t^4/3)$
(TCL4). It is interesting to note that the second order expansions of the
TCL and NZ master equations exhibit a $v_x\leftrightarrow v_y$ symmetry
between the components of the Bloch vector, and only the third order
correction breaks this symmetry. Notice that the coefficient of $\alpha^3$
does not vanish in this model unlike in the one considered in \cite{BBP04}
because both $\left\la\mathcal{L}^3\right\ra_{pc}\neq 0$ and $\left\la%
\mathcal{L}^3\right\ra_{oc}\neq 0$ and hence the third order (and other odd
order) approximations exist.

\subsection{Post-Markovian (PM) master equation}

In this section we study the performance of the post-Markovian master
equation recently proposed in \cite{ShabaniLidar:05}: 
\begin{equation}
\frac{\partial \rho (t)}{\partial t}=\mathcal{D}\int_{0}^{t}dt^{\prime
}k(t^{\prime })\exp (\mathcal{D}t^{\prime })\rho (t-t^{\prime })\mathrm{.}
\label{PMAL}
\end{equation}%
This equation was constructed via an interpolation between the exact
dynamics and the dynamics in the Markovian limit. The operator $\mathcal{D}$
is the dissipator in the Lindblad equation (\ref{Lindblad}), and $k(t)$ is a
phenomenological memory kernel which must be found by fitting to data or
guessed on physical grounds. As was discussed earlier, the Markovian
approximation fails for our model, nevertheless, one can use the form of the
dissipator we obtained in Eq. (\ref{Lindblad}) 
\begin{equation}
\mathcal{D}\rho =\sigma ^{z}\rho \sigma ^{z}-\rho .  \label{dissipator}
\end{equation}%
It is interesting to examine to what extent Eq. (\ref{PMAL}) can approximate
the exact dynamics. As a measure of the performance of the post-Markovian
equation, we will take the trace-distance between the exact solution $\rho _{%
\mathrm{exact}}(t)$ and the solution to the post-Markovian equation $\rho
_{1}(t)$. The general solution of Eq. (\ref{PMAL}) can be found by
expressing $\rho (t)$ in the damping basis \cite{Briegel:93} and applying a
Laplace transform \cite{ShabaniLidar:05}. The solution is%
\begin{equation}
\rho (t)=\sum_{i}\mu _{i}(t)R_{i}=\sum_{i}{\text{Tr}}(L_{i}\rho (t))R_{i},
\end{equation}%
where 
\begin{equation}
\mu _{i}(t)=\mathrm{Lap}^{-1}\left[ \frac{1}{s-\lambda _{i}\tilde{k}%
(s-\lambda _{i})}\right] \mu _{i}(0)\equiv \xi _{i}(t)\mu _{i}(0),
\end{equation}%
($\mathrm{Lap}^{-1}$ is the inverse Laplace transform) with $\tilde{k}$
being the Laplace transform of the kernel $k$, $\{L_{i}\}$ and $\{R_{i}\}$
being the left and right eigenvectors of the superoperator $\mathcal{D}$,
and $\lambda _{i}$ the corresponding eigenvalues. For our dissipator the
damping basis is $\{L_{i}\}=\{R_{i}\}=\{\frac{I}{\sqrt{2}},\frac{\sigma ^{x}%
}{\sqrt{2}},\frac{\sigma ^{y}}{\sqrt{2}},\frac{\sigma ^{z}}{\sqrt{2}}\}$ and
the eigenvalues are $\{0,-2,-2,0\}$. Therefore, we can immediately write the
formal solution in terms of the Bloch vector components: 
\begin{gather}
v_{x,y}(t)={\text{Lap}}^{-1}\left[ \frac{1}{s+2\tilde{k}(s+2)}\right]
v_{x,y}(0)\equiv \xi (t)v_{x,y}(0).  \label{eq:vy-post}
\end{gather}

We see that $v_{x}(t)$ has no dependence on $v_{y}(0)$, and neither does $%
v_{y}(t)$ on $v_{x}(0)$, in contrast to the exact solution. The difference
comes from the fact that the dissipator $\mathcal{D}$ does not couple $%
v_{x}(t)$ and $v_{y}(t)$. This reveals an inherent limitation of the
post-Markovian master equation:\ it inherits the symmetries of the Markovian
dissipator $\mathcal{D}$, which may differ from those of the generator of
the exact dynamics. In order to rigorously determine the optimal
performance, we use the trace distance between the exact solution and a
solution to the post-Markovian equation:

\begin{eqnarray}
D(\rho _{\mathrm{exact}}(t),\rho _{1}(t)) &=&\frac{1}{2}\sqrt{(C(t)-\xi
(t))^{2}+S(t)^{2}}  \notag \\
&\times &\sqrt{v_{x}(0)^{2}+v_{y}(0)^{2}}.
\end{eqnarray}%
Obviously this quantity reaches its minimum for $\xi (t)=C(t),\forall t$
independently of the initial conditions. The kernel for which the optimal
performance of the post-Markovian master equation is achieved, can thus be
formally expressed, using Eq. (\ref{eq:vy-post}), as: 
\begin{equation}
k_{\mathrm{opt}}(t)=\frac{1}{2}e^{2t}\mathrm{Lap}^{-1}\left\{ \frac{1}{%
\mathrm{Lap}(C(t))}-s\right\} .  \label{kopt}
\end{equation}%
It should be noted that the condition for complete positivity of the map
generated by Eq. (\ref{PMAL}), $\sum_{i}\xi _{i}(t)L_{i}^{T}\otimes
R_{i}\geq 0$ \cite{ShabaniLidar:05}, amounts here to $|\xi (t)|=|C(t)|\leq 1$%
, which holds for all $t$. Thus the minimum achievable trace-distance
between the two solutions is given by 
\begin{equation}
D_{\mathrm{min}}(\rho _{\mathrm{exact}}(t),\rho _{1}(t))=\frac{1}{2}S(t)%
\sqrt{v_{x}(0)^{2}+v_{y}(0)^{2}}.
\end{equation}%
The optimal fit is plotted in Sec. IV.

Finding a simple analytical expression for the optimal kernel Eq. (\ref{kopt}%
) seems difficult due to the complicated form of $C(t)$. One way to approach
this problem is to expand $C(t)$ in powers of $\alpha t$ and consider terms
which give a valid approximation for small times $\alpha t\ll 1$. For
example, Eq. (\ref{eq:f-approx}) yields the lowest non-trivial order as: 
\begin{equation}
C_{2}(t)=1-2Q_{2}\alpha ^{2}t^{2}+\mathcal{O}(\alpha ^{4}t^{4}).
\end{equation}%
Note that this solution violates the complete positivity condition for times
larger than $t=1/\alpha \sqrt{2Q_{2}}$. The corresponding kernel is: 
\begin{equation}
k_{2}(t)=2\alpha ^{2}Q_{2}e^{2t}\cosh (2\sqrt{Q_{2}}\alpha t).
\end{equation}%
Alternatively we could try finding a kernel that matches some of the
approximate solutions discussed so far. For example, it turns out that the
kernel 
\begin{equation}
k_{\mathrm{NZ2}}(t)=2\alpha ^{2}Q_{2}e^{2t}
\end{equation}%
leads to an exact match of the NZ2 solution. Finding a kernel which gives a
good description of the evolution of an open system is an important but in
general, difficult question which remains open for further investigation. We
note that this question was also taken up in the context of the PM\ in the
recent study \cite{ManPet06}, where the PM\ was applied to an exactly
solvable model describing a qubit undergoing spontaneous emission and
stimulated absorption. No attempt was made to optimize the memory kernel and
hence the agreement with the exact solution was not as impressive as might
be possible with optimization.

\section{Comparison of the analytical solution and the different
approximation techniques}

In the results shown below, all figures express the evolution in terms of
the dimensionless parameter $\alpha t$ (plotted on a logarithmic scale). We
choose the initial condition $v_{x}(0)=v_{y}(0)=1/\sqrt{2}$ and plot only $%
v_{x}(t)$ since the structure of the equations for $v_{x}(t)$ and $v_{y}(t)$
is similar. In order to compare the different methods of approximation, we
consider various choices of parameter values in our model.

\subsection{Exact Solution}

\begin{figure}[h]
\begin{center}
\includegraphics[scale=0.3]{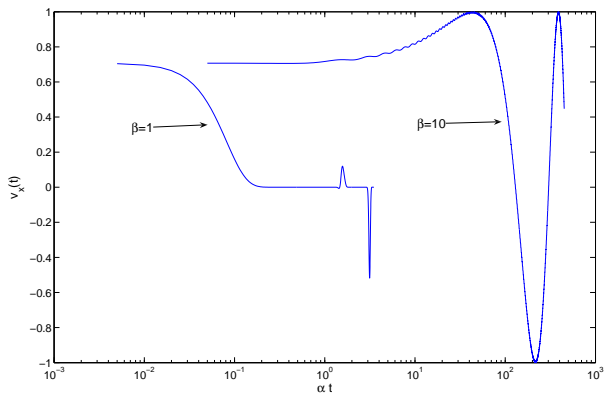}
\end{center}
\caption{(Color online) Comparison of the exact solution at $\protect\beta =1
$ and $\protect\beta =10$ for $N=100$.}
\label{N100_exact}
\end{figure}

\begin{figure}[h]
\begin{center}
\includegraphics[scale=0.3]{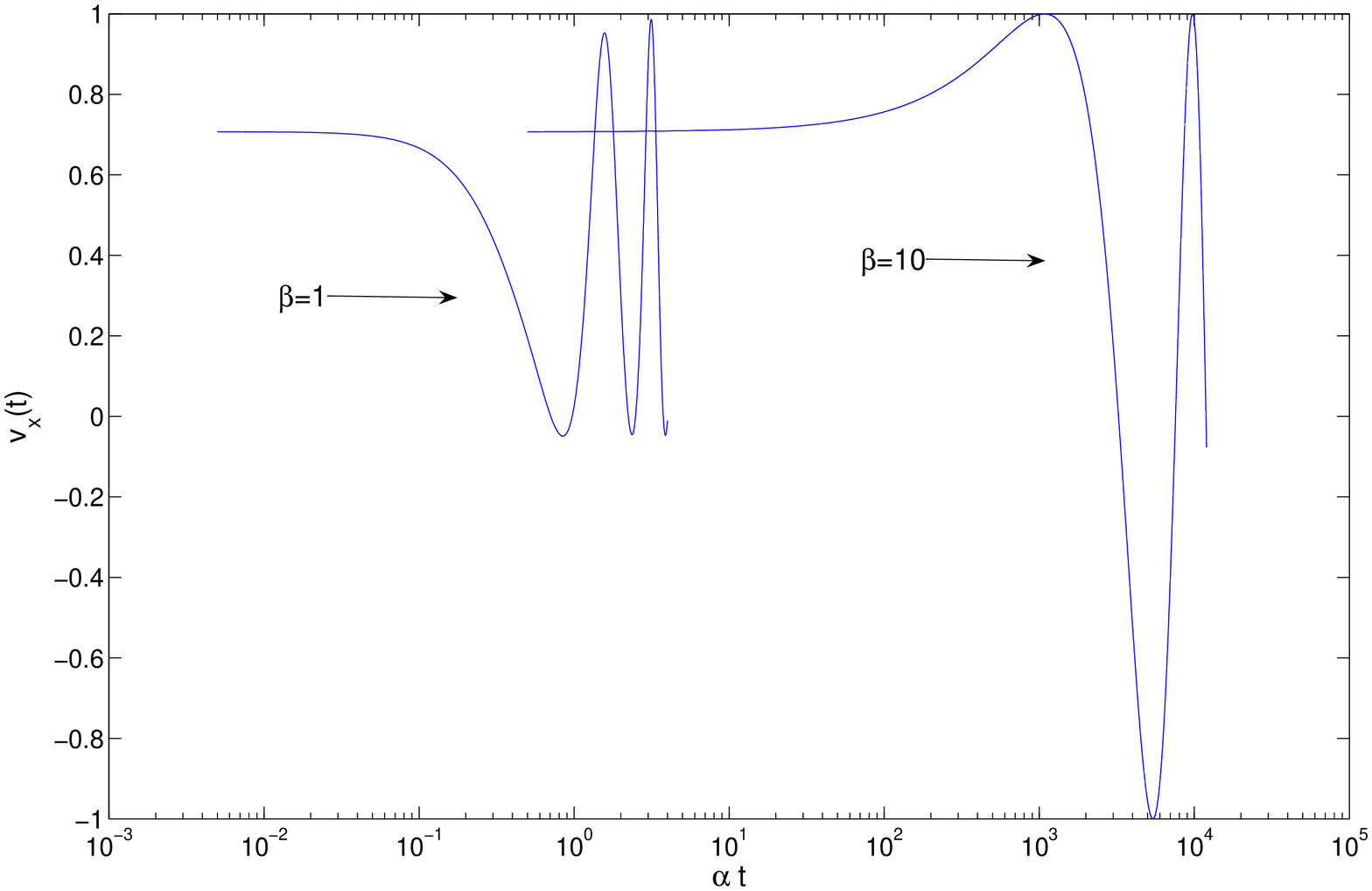}
\end{center}
\caption{(Color online) Comparison of the exact solution at $\protect\beta =1
$ and $\protect\beta =10$ for $N=4$.}
\label{N4_exact}
\end{figure}

We first assume that the frequencies of the qubits in the bath are equal ($%
\Omega _{n}=1$, $\forall n$), and so are the coupling constants ($g_{n}=1$, $%
\forall n$). In this regime, we consider large and small numbers of bath
spins $N=100$ and $N=4$, and two different temperatures $\beta =1$ and $%
\beta =10$. Figs.~\ref{N100_exact} and \ref{N4_exact} show the exact
solution for $N=100$ and $N=4$ spins, respectively, up to the second
recurrence time. For each $N$, we plot the exact solution for $\beta =1$ and 
$\beta =10$.

\begin{figure}[h]
\begin{center}
\includegraphics[scale=0.3]{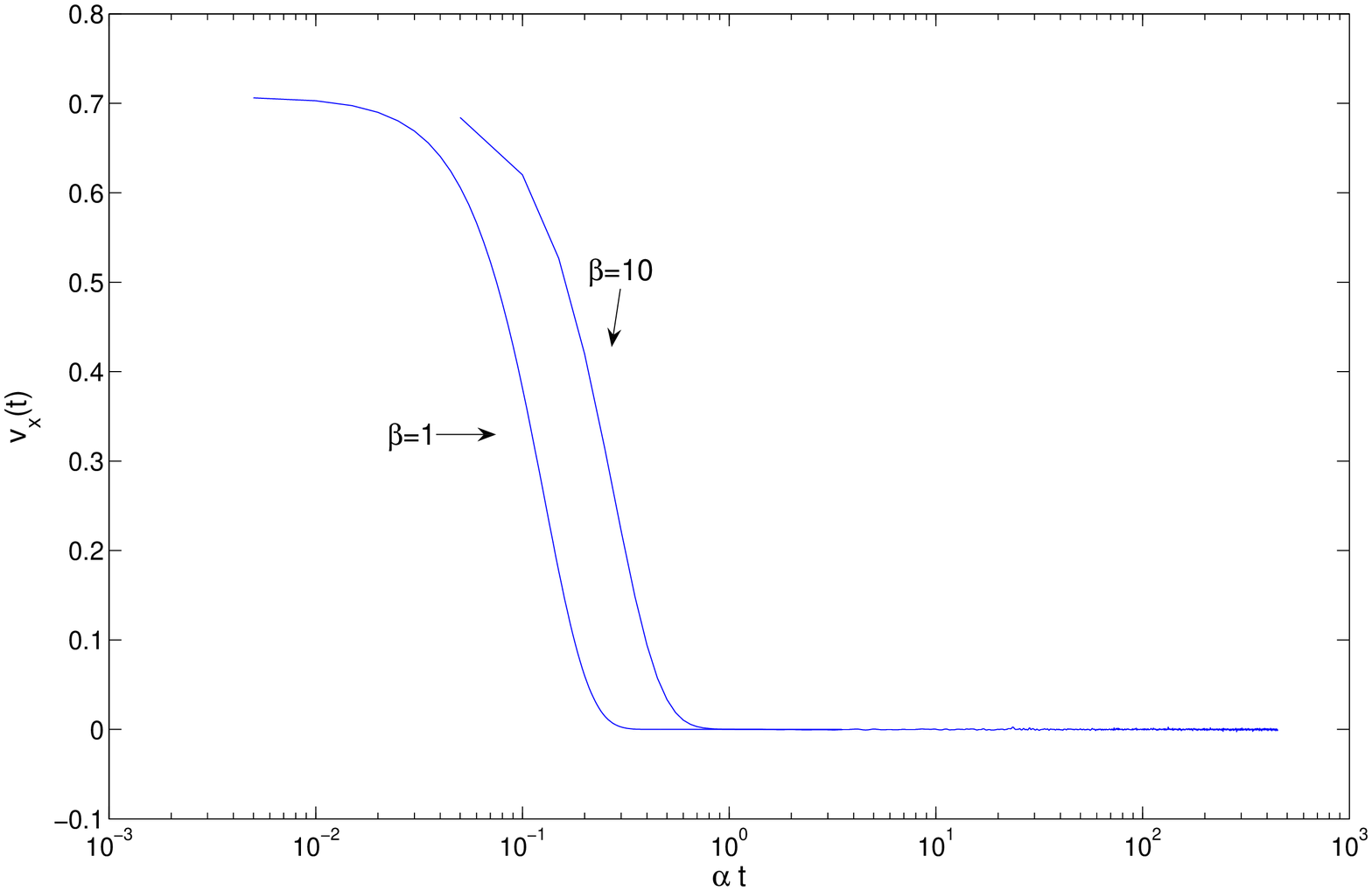}
\end{center}
\caption{(Color online) Comparison of the exact solution at $\protect\beta =1
$ and $\protect\beta =10$ for $N=100$ for randomly generated $g_{n}$ and $%
\Omega _{n}$.}
\label{N100_exact_random_long}
\end{figure}

We also consider the case where the frequencies $\Omega _{n}$ and the
coupling constants $g_{n}$ can take different values. We generated uniformly
distributed random values in the interval $[-1,1]$ for both $\Omega _{n}$
and $g_{n}$. In Figs.~(\ref{N100_exact_random_long}) and (\ref%
{N4_exact_random}) we plot the ensemble average of the solution over 50
random ensembles. The main difference from the solution with equal $\Omega
_{n}$ and $g_{n}$ is that the partial recurrences decrease in size,
especially as $N$ increases. We attribute this damping partially to the fact
that we look at the ensemble average, which amounts to averaging out the
positive and negative oscillations that arise for different values of the
parameters. The main reason, however, is that for a generic ensemble of
random $\Omega _{n}$ and $g_{n}$ the positive and negative oscillations in
the sums \eqref{CSeqn} tend to average out. This is particularly true for
large $N$, as reflected in Fig.~\ref{N100_exact_random_long}. We looked at a
few individual random cases for $N=100$ and recurrences were not present
there. For $N=20$ (not shown here), some small recurrences were still
visible.

We also looked at the case where one of the coupling constants, say $g_i$,
has a much larger magnitude than the other ones (which were made equal). The
behavior was similar to that for a bath consisting of only a single spin.

\begin{figure}[h]
\begin{center}
\includegraphics[scale=0.3]{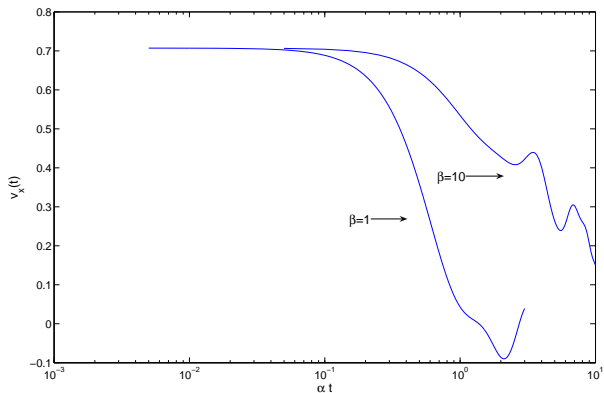}
\end{center}
\caption{(Color online) Comparison of the exact solution at $\protect\beta =1
$ and $\protect\beta =10$ for $N=4$ for randomly generated $g_{n}$ and $%
\Omega _{n}$.}
\label{N4_exact_random}
\end{figure}

In the following, we plot the solutions of different orders of the NZ, TCL
and PM master equations and compare them for the same parameter values.

\subsection{NZ}

\begin{figure}[h]
\begin{center}
\includegraphics[scale=0.3]{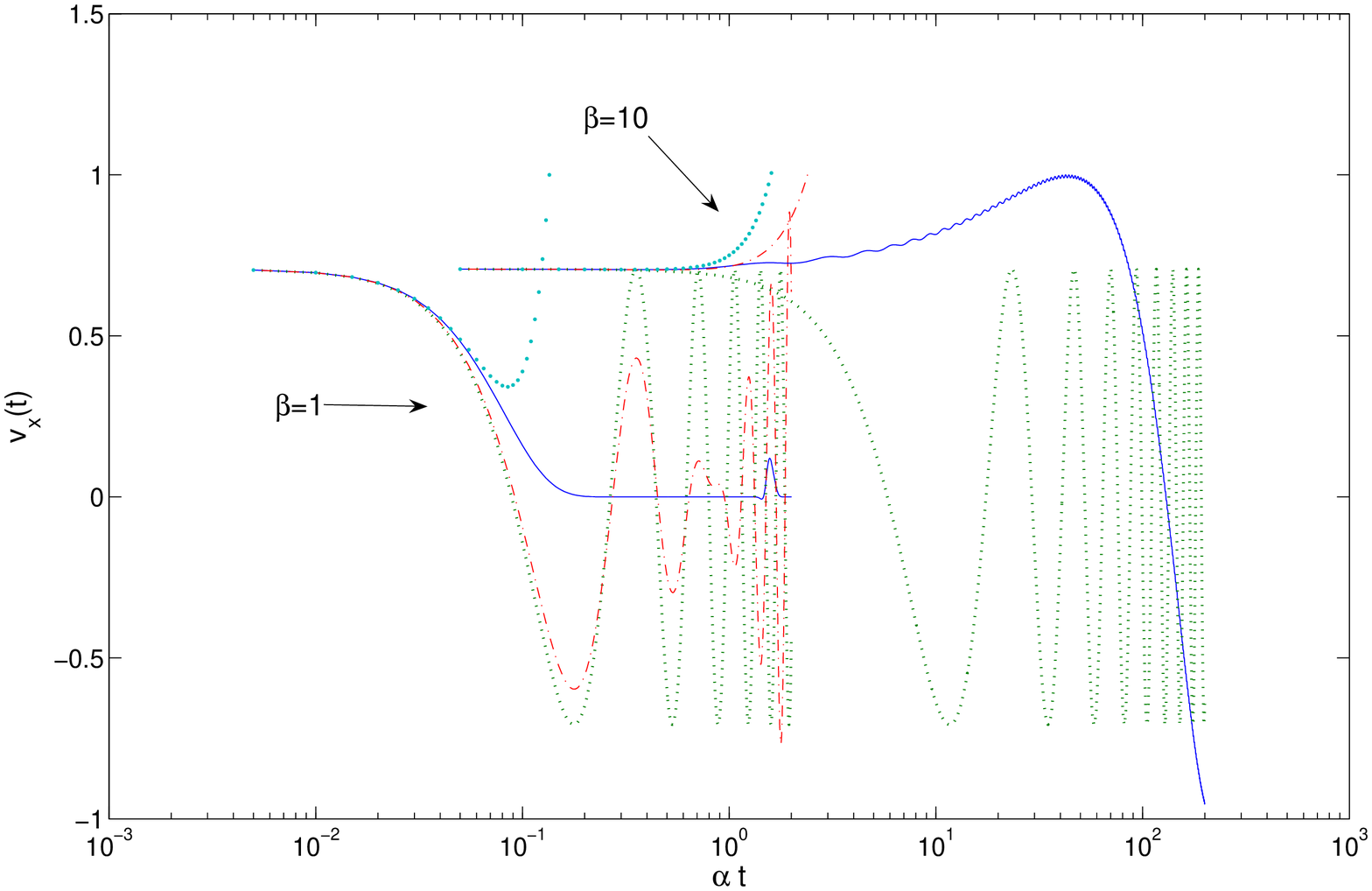}
\end{center}
\caption{(Color online) Comparison of the exact solution, NZ2, NZ3 and NZ4
at $\protect\beta =1$ and $\protect\beta =10$ for $N=100$. The exact
solution is the solid (blue) line, NZ2 is the dashed (green) line, NZ3 is
the dot-dashed (red) line and NZ4 is the dotted (cyan) line.}
\label{N100_NZ}
\end{figure}

\begin{figure}[h]
\begin{center}
\includegraphics[scale=0.3]{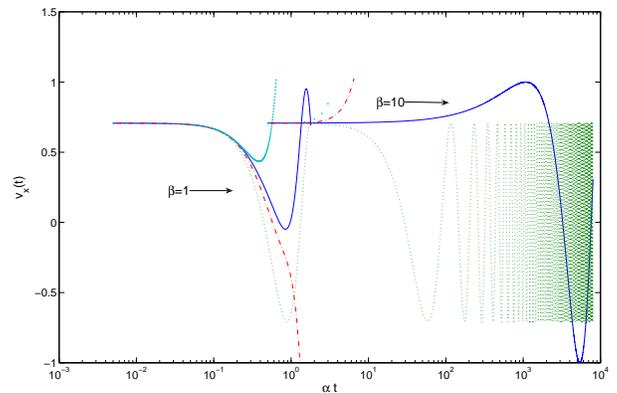}
\end{center}
\caption{(Color online) Comparison of the exact solution, NZ2, NZ3 and NZ4
at $\protect\beta =1$ and $\protect\beta =10$ for $N=4$. The exact solution
is the solid (blue) line, NZ2 is the dashed (green) line, NZ3 is the
dot-dashed (red) line and NZ4 is the dotted (cyan) line.}
\label{N4_NZ}
\end{figure}

In this subsection, we compare the solutions of different orders of the NZ
master equation for $\Omega _{n}=g_{n}=1$. Fig.~(\ref{N100_NZ}) shows the
solutions to NZ2, NZ3, NZ4 and the exact solution for $\beta =1$ and $\beta
=10$ up to the first recurrence time of the exact solution. For short times
NZ4 is the better approximation. It can be seen that while NZ2 and NZ3 are
bounded, NZ4 leaves the Bloch sphere. But note that the approximations under
which these solutions have been obtained are valid for $\alpha t\ll 1$. The
NZ4 solution leaves the Bloch sphere in a regime where the approximation is
not valid. For $\beta =10$, NZ2 again has a periodic behavior (which is
consistent with the solution), while the NZ3 and NZ4 solutions leave the
Bloch sphere after small times. Fig.~(\ref{N4_NZ}) shows the same graphs for 
$N=4$. In this case both NZ3 and NZ4 leave the Bloch sphere for $\beta =1$
and $\beta =10$, while NZ2 has a periodic behavior. A clear conclusion from
these plots is that the NZ\ approximation is truly a short-time one:\ it
becomes completely unreliable for times longer than $\alpha t\ll 1$.

\subsection{TCL}

\begin{figure}[h]
\begin{center}
\includegraphics[scale=0.3]{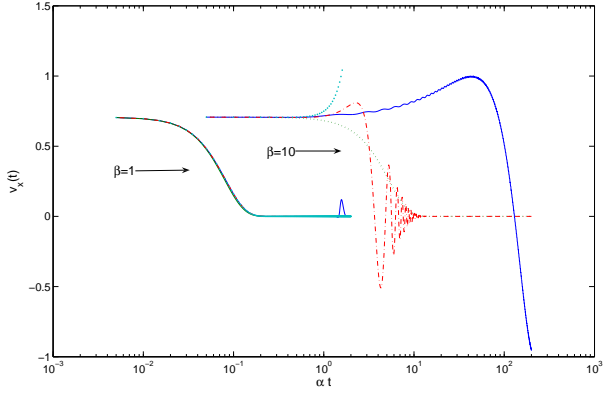}
\end{center}
\caption{(Color online) Comparison of the exact solution, TCL2, TCL3 and
TCL4 at $\protect\beta =1$ and $\protect\beta =10$ for $N=100$. The exact
solution is the solid (blue) line, TCL2 is the dashed (green) line, TCL3 is
the dot-dashed (red) line and TCL4 is the dotted (cyan) line. Note that for $%
\protect\beta =1$, the curves nearly coincide.}
\label{N100_TCL}
\end{figure}

\begin{figure}[h]
\begin{center}
\includegraphics[scale=0.3]{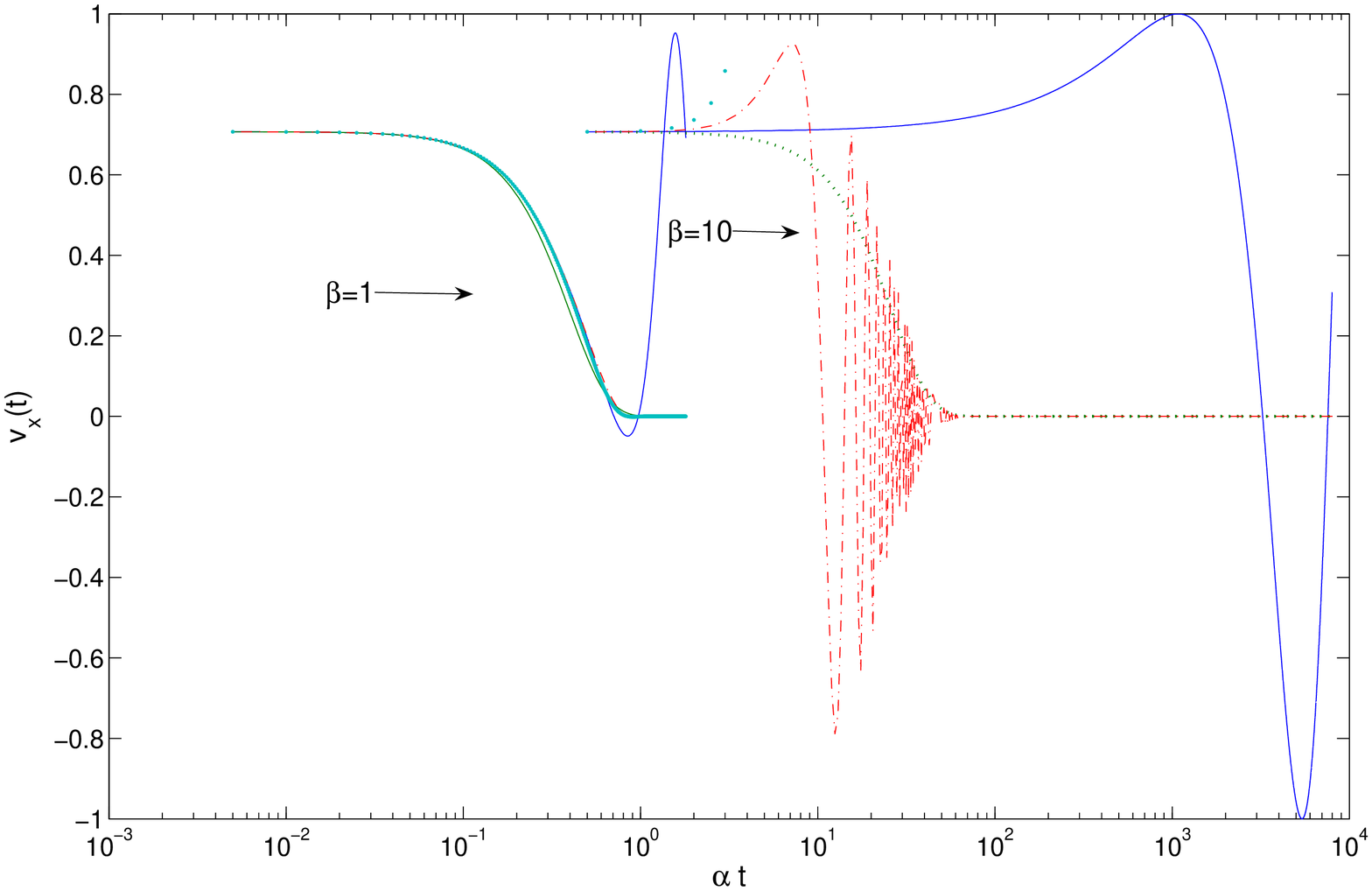}
\end{center}
\caption{(Color online) Comparison of the exact solution, TCL2, TCL3 and
TCL4 at $\protect\beta =1$ and $\protect\beta =10$ for $N=4$. The exact
solution is the solid (blue) line, TCL2 is the dashed (green) line, TCL3 is
the dot-dashed (red) line and TCL4 is the dotted (cyan) line. Note that for $%
\protect\beta =1$, TCL3, TCL4 and the exact solution nearly coincide.}
\label{N4_TCL}
\end{figure}

Fig.~(\ref{N100_TCL}) plots the exact solution, TCL2, TCL3 and TCL4 at $%
\beta =1$ and $\beta =10$ for $N=100$ spins and $\Omega _{n}=g_{n}=1$. It
can be seen that for $\beta =1$, the TCL solution approximates the exact
solution well even for long times. However, the TCL solution cannot
reproduce the recurrence behavior of the exact solution (also shown in the
figure.) Fig.~(\ref{N4_TCL}) shows the same graphs for $N=4$. In this case,
while TCL2 and TCL3 decay, TCL4 increases exponentially and leaves the Bloch
sphere after a short time. This is because the exponent in the solution of
TCL4 in Eq.~(\ref{TCL3}) is positive. Here again the approximations under
which the solutions have been obtained are valid only for small time scales
and the graphs demonstrate the complete breakdown of the perturbation
expansion for large values of $\alpha t$. Moreover, the graphs reveal the
sensitivity of the approximation to temperature:\ the TCL\ fares much better
at high temperatures.

\begin{figure}[h]
\begin{center}
\includegraphics[scale=0.3]{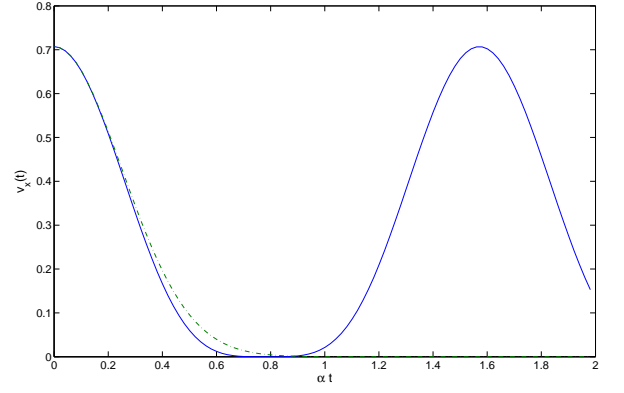}
\end{center}
\caption{(Color online) Comparison of TCL2 and the exact solution to
demonstrate the validity of the TCL approximation for $N=4$ and $\protect%
\beta =1$. The solid (blue) line denotes the exact solution and the dashed
(green) line is TCL2. Note that the time axis here is on a linear scale.
TCL2 breaks down at $\protect\alpha t\approx 0.9$, where it remains flat,
while the exact solution has a recurrence.}
\label{N4B0TCL2}
\end{figure}

In order to determine the validity of the TCL approximation, we look at the
invertibility of the Kraus map derived in Eq. (\ref{KrausForm}) or
equivalently Eq. (\ref{CSeqn}). As mentioned earlier, this map is
non-invertible if $C(t)^{2}+S(t)^{2}=0$ for some $t$ (or equivalently $%
v_{x}(t)=0$ and $v_{y}(t)=0$). This will happen if and only if at least one
of the $\beta _{n}$ is zero. This can occur when the bath density matrices
of some of the bath spins are maximally mixed or in the limit of a very high
bath temperature. Clearly, when the Kraus map is non-invertible, the TCL
approach becomes invalid since it relies on the assumption that the
information about the initial state is contained in the current state. This
fact has also been observed for the spin-boson model with a damped
Jaynes-Cummings Hamiltonian \cite{Breuer:book}. At the point where the Kraus
map becomes non-invertible, the TCL solution deviates from the exact
solution (see Fig.~ \ref{N4B0TCL2}). We verified that both $v_{x}$ and $v_{y}
$ vanish at this point.

\subsection{NZ, TCL, and PM}

\begin{figure}[h]
\begin{center}
\includegraphics[scale=0.3]{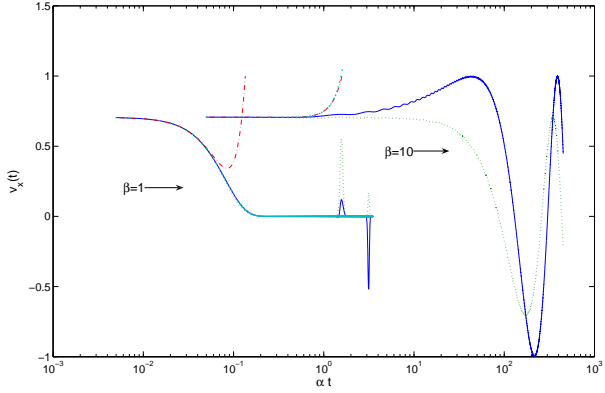}
\end{center}
\caption{(Color online) Comparison of the exact solution, NZ4, TCL4 and PM
at $\protect\beta =1$ and $\protect\beta =10$ for $N=100$. The exact
solution is the solid (blue) line, PM is the dashed (green) line, NZ4 is the
dot-dashed (red) line and TCL4 is the dotted (cyan) line. Note that for $%
\protect\beta =1$, TCL4, PM and the exact solution nearly coincide for short
and medium times. Only PM\ captures the recurrences of the exact solution at
long times.}
\label{N100_best}
\end{figure}

In this subsection, we compare the exact solution to TCL4, NZ4 and the
solution of the optimal PM master equation. Fig.~(\ref{N100_best}) shows
these solutions for $N=100$ and $\beta =1$ and $\beta =10$ when $\Omega
_{n}=g_{n}=1$. Here we observe that while the short-time behavior of the
exact solution is approximated well by all the approximations we consider,
the long-time behavior is approximated well only by PM. 

\begin{figure}[h]
\begin{center}
\includegraphics[scale=0.3]{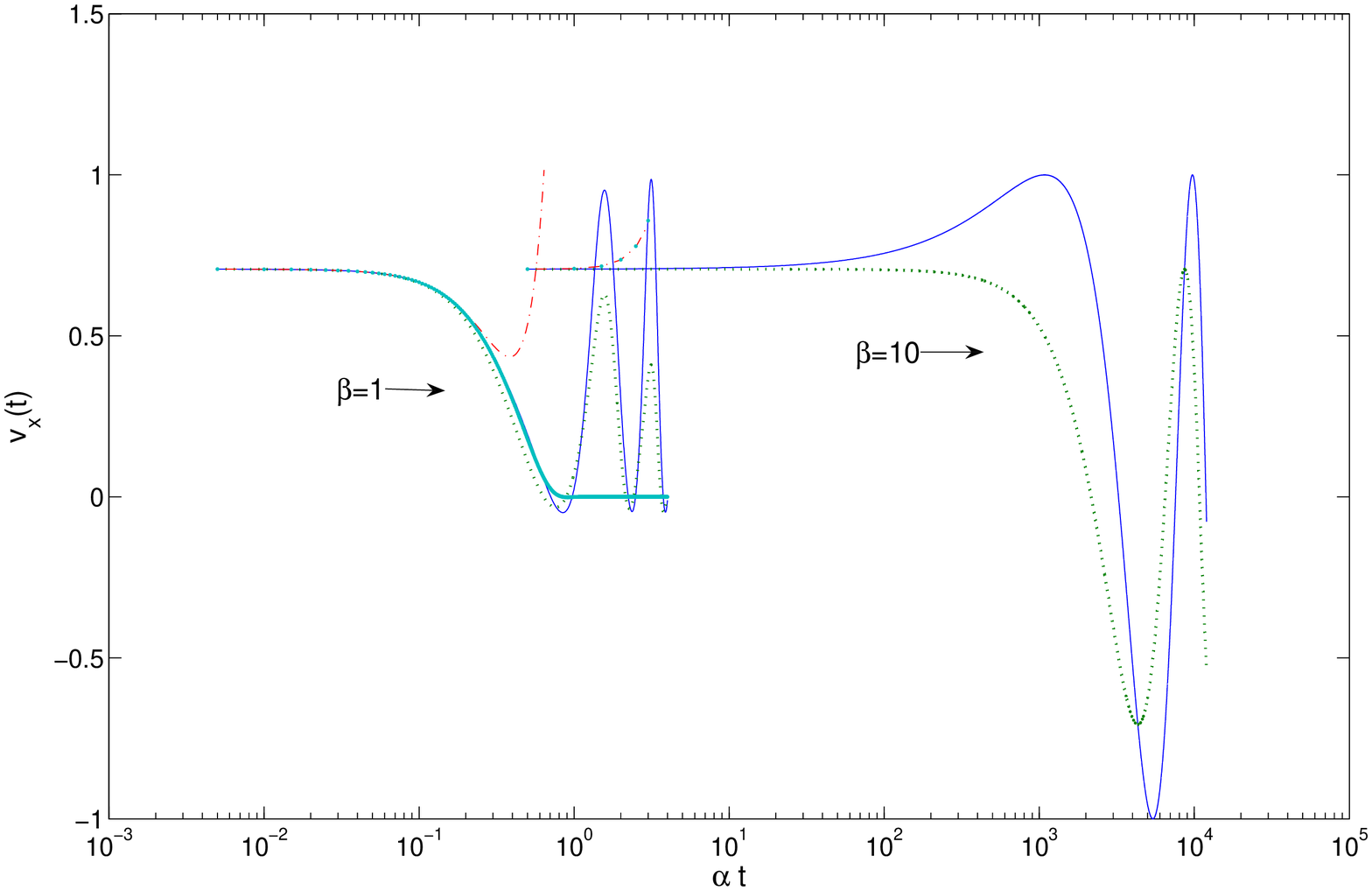}
\end{center}
\caption{(Color online) Comparison of the exact solution, NZ4, TCL4 and PM
at $\protect\beta=1$ and $\protect\beta=10$ for $N=4$. The exact solution is
the solid (blue) line, PM is the dashed (green) line, NZ4 is the dot-dashed
(red) line and TCL4 is the dotted (cyan) line. Note that for $\protect\beta=1
$, TCL4 and the exact solution nearly coincide for short and medium times.}
\label{N4_best}
\end{figure}

\begin{figure}[h]
\begin{center}
\includegraphics[scale=0.3]{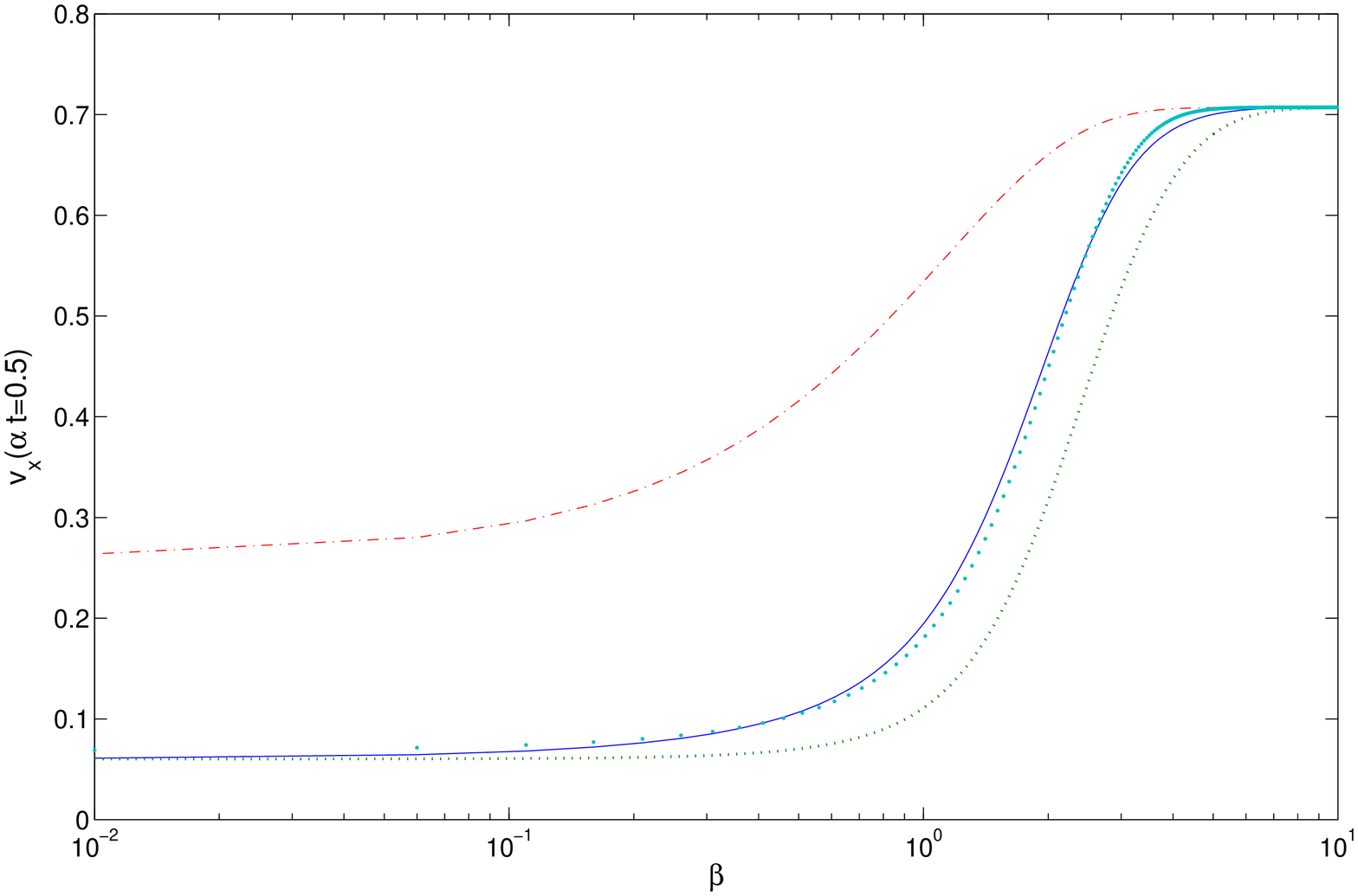}
\end{center}
\caption{(Color online) Comparison of the exact solution, NZ4, TCL4 and PM
at $\protect\alpha t=0.5$ for $N=4$ for different $\protect\beta\in [0.01,10]
$. The exact solution is the solid (blue) line, PM is the dashed (green)
line, NZ4 is the dot-dashed (red) line and TCL4 is the dotted (cyan) line.}
\label{N4_beta}
\end{figure}

For $\beta =1$, NZ4 leaves the Bloch sphere after a short time while TCL4
decays with the exact solution. But as before, the TCL solution cannot
reproduce the recurrences seen in the exact solution. The optimal PM
solution, by contrast, is capable of reproducing both the decay and the
recurrences. TCL4 and NZ4 leave the Bloch sphere after a short time for $%
\beta =10$, while PM again reproduces the recurrences in the exact solution.
Fig.~\ref{N4_best} shows the corresponding graphs for $N=4$ and it can be
seen that again PM can outperform both TCL and NZ for long times. Figs.~\ref%
{N100_beta} and \ref{N4_beta} show the performance of TCL4, NZ4 and PM
compared to the exact solution at a fixed time (for which the approximations
are valid) for different temperatures ($\beta \in \lbrack 0.01,10]$). It can
be seen that both TCL4 and the optimal PM solution perform better than NZ4
at medium and high temperatures, with TCL4 outperforming PM at medium
temperatures. The performance of NZ4 is enhanced at low temperatures, where
it performs similarly to TCL4 (see also Figs.~\ref{N100_best} and \ref%
{N4_best}). This can be understood from the short-time approximation to the
exact solution given in Eq. (\ref{shorttimes}), which up to the precision
for which it was derived is also an approximation of NZ2 [Eq. (\ref{Born})].
As discussed above, this approximation (which also coincides with TCL2) is
valid when $2Q_{2}(\alpha t)^{2}\ll 1$. As temperature decreases, so does
the magnitude of $Q_{2}$, which leads to a better approximation at fixed $%
\alpha t$. Since NZ2 gives the lowest-order correction, this improvement is
reflected in NZ4 as well.

\begin{figure}[h]
\begin{center}
\includegraphics[scale=0.3]{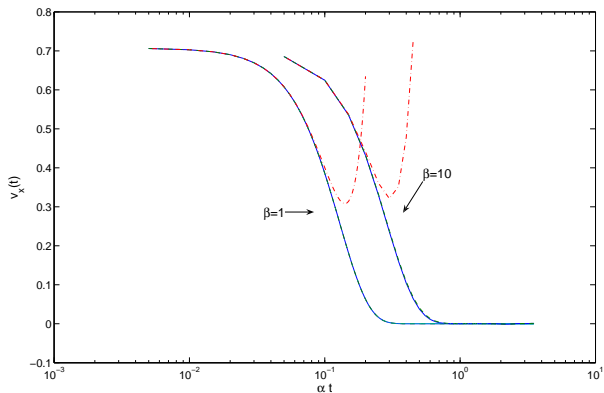}
\end{center}
\caption{(Color online) Comparison of the exact solution, NZ4, TCL4 and PM
at $\protect\beta=1$ and $\protect\beta=10$ for $N=100$ for random values of 
$g_n$ and $\Omega_n$. The exact solution is the solid (blue) line, PM is the
dashed (green) line, NZ4 is the dot-dashed (red) line and TCL4 is the dotted
(cyan) line. Note that for $\protect\beta=1$ and $\protect\beta=10$, TCL4,
PM and the exact solution nearly coincide.}
\label{N100_best_random}
\end{figure}

\begin{figure}[h]
\begin{center}
\includegraphics[scale=0.3]{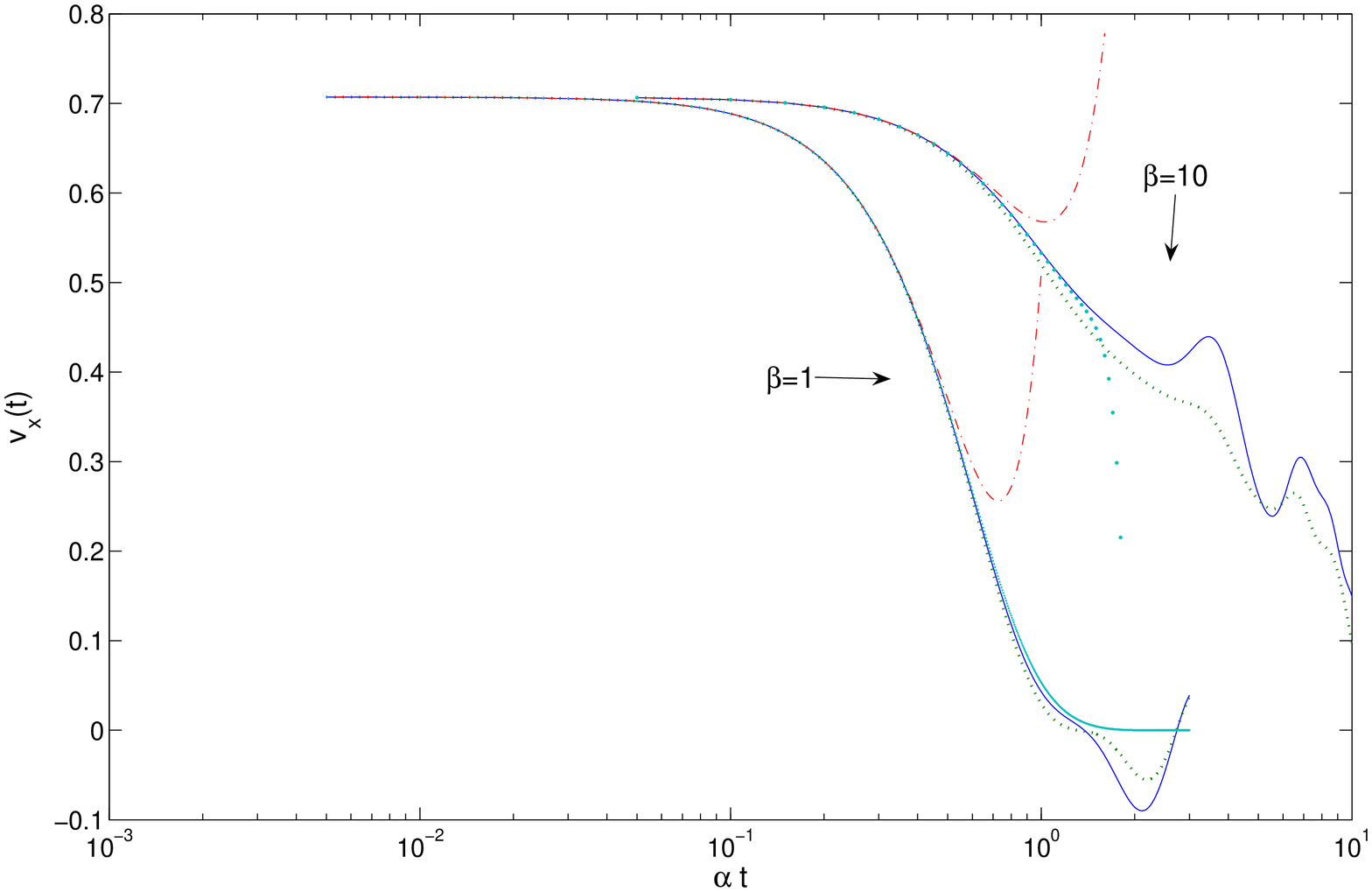}
\end{center}
\caption{(Color online) Comparison of the exact solution, NZ4, TCL4 and PM
at $\protect\beta=1$ and $\protect\beta=10$ for $N=4$ for random values of $%
g_n$ and $\Omega_n$. The exact solution is the solid (blue) line, PM is the
dashed (green) line, NZ4 is the dot-dashed (red) line and TCL4 is the dotted
(cyan) line. Note that for $\protect\beta=1$, TCL4, PM and the exact
solution nearly coincide for short and medium times.}
\label{N4_best_random}
\end{figure}

In Figs. \ref{N100_best_random} and \ref{N4_best_random} we plot the
averaged solutions over 50 ensembles of random values for $\Omega _{n}$ and $%
g_{n}$ in the interval $[-1,1]$. We see that on average TCL4, NZ4 and the
optimal PM solution behave similarly to the case when $\Omega _{n}=g_{n}=1$.
Due to the damping of the recurrences, especially when $N=100$, the TCL4 and
the PM solutions match the exact solution closely for much longer times than
in the deterministic case. Again, the PM solution is capable of
qualitatively matching the behavior of the exact solution at long times.

\begin{figure}[h]
\begin{center}
\includegraphics[scale=0.3]{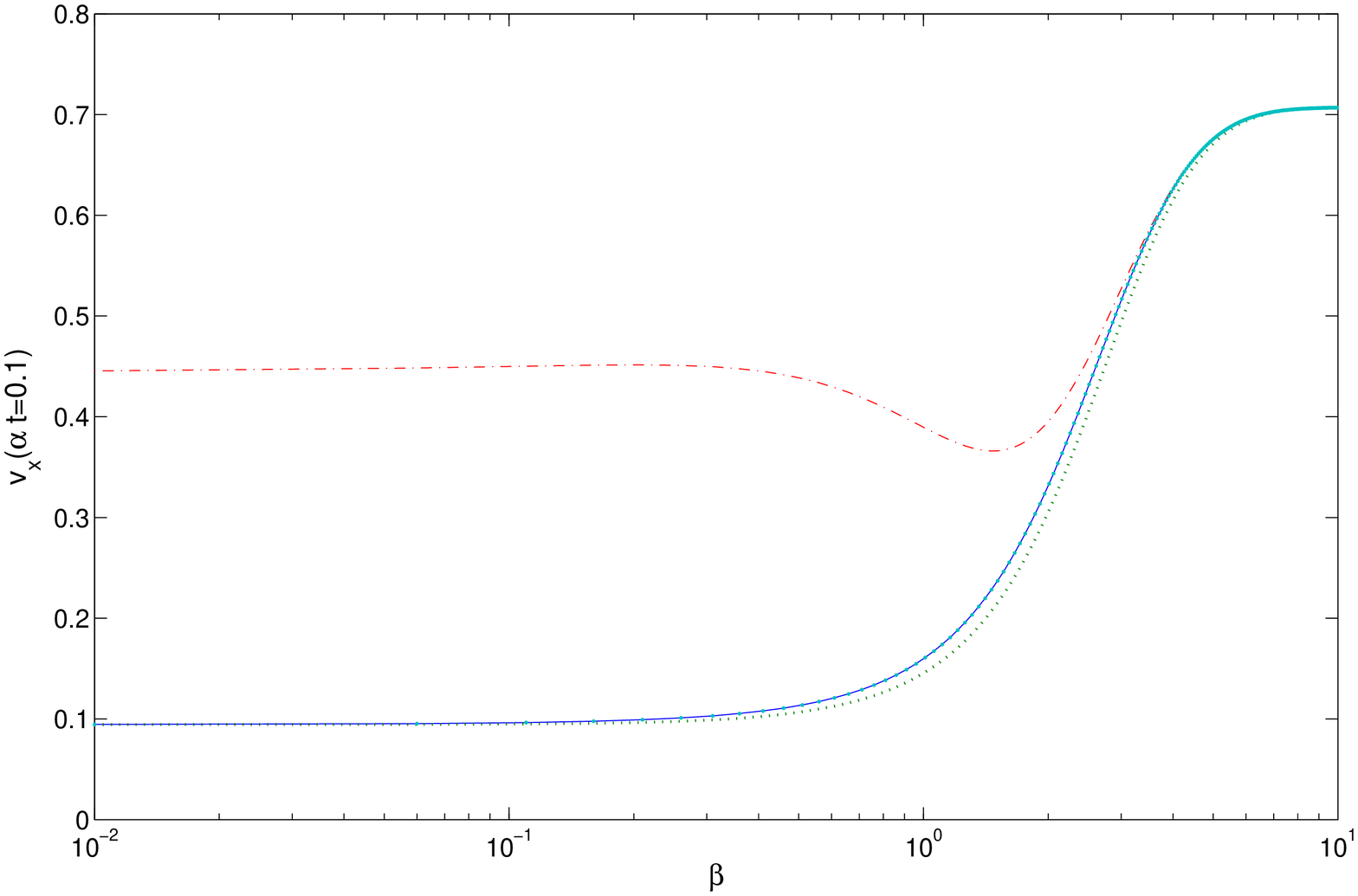}
\end{center}
\caption{(Color online) Comparison of the exact solution, NZ4, TCL4 and PM
at $\protect\alpha t=0.1$ for $N=100$ for different $\protect\beta\in
[0.01,10]$. The exact solution is the solid (blue) line, PM is the dashed
(green) line, NZ4 is the dot-dashed (red) line and TCL4 is the dotted (cyan)
line.}
\label{N100_beta}
\end{figure}

\subsection{Coarse-graining approximation}

\begin{figure}[h]
\begin{center}
\includegraphics[scale=0.3]{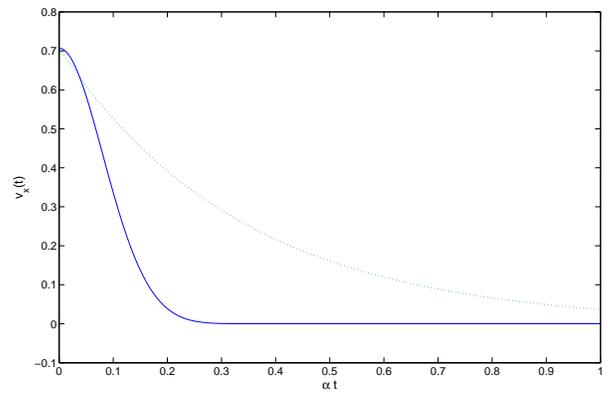}
\end{center}
\caption{(Color online) Comparison of the exact solution and the optimal
coarse-graining approximation for $N=50$ and $\protect\beta =1$. The exact
solution is the solid (blue) line and the coarse-graining approximation is
the dashed (green) line. Note the linear scale time axis.}
\label{N50B1CG}
\end{figure}

Finally, we examine the coarse-graining approximation discussed in Sec. III.
We choose the time over which the average trace distance is calculated to be
the time where the exact solution dies down. In Fig. \ref{N50B1CG} we plot
the coarse-grained solution for the value of $\tau $ for which the trace
distance to the exact solution is minimum. As can be seen, the
coarse-graining approximation does not help since the Markovian assumption
is not valid for this model. In deriving the coarse-graining approximation 
\cite{Lidar:CP01} one makes the assumption that the coarse-graining time
scale is greater than any characteristic bath time scale. But the
characteristic time scale of the bath is infinite in this case.

\section{Summary and Conclusions}

We studied the performance of various methods for approximating the
evolution of an Ising model of an open quantum system for a qubit system
coupled to a bath bath consisting of $N$ qubits. The high symmetry of the
model allowed us to derive the exact dynamics of the system as well as find
analytical solutions for the different master equations. We saw that the
Markovian approximation fails for this model due to the time independence of
the bath correlation functions. This is also reflected in the fact that the
coarse-graining method \cite{Lidar:CP01} does not approximate the exact
solution well. We discussed the performance of these solutions for various
parameter regimes. Unlike other spin bath models discussed in literature
(e.g., \cite{BBP04}), the odd-order bath correlation functions do not
vanish, leading to the existence of odd-order terms in the solution of TCL
and NZ equations. These terms describe the rotation around the $z$ axis of
the Bloch sphere, a fact which is reflected in the exact solution. We showed
that up to fourth order TCL performs better than NZ at medium and high
temperatures. For low temperatures we demonstrated an enhancement in the
performance of NZ and showed that NZ and TCL perform equally well. We showed
that the TCL approach breaks down for certain parameter choices and related
this to the non-invertibility of the Kraus map describing the system
dynamics. We also studied the performance of the post-Markovian master
equation obtained in \cite{ShabaniLidar:05} with an optimal memory kernel.
We discussed possible ways of approximating the optimal kernel for short
times and derived the kernel which leads to an exact fit to the NZ2
solution. It turns out that PM master equation performs as well as the TCL2
for a large number of spins and outperforms all orders of NZ\ and TCL\
considered here at long times, as it captures the recurrences of the exact
solution. 

Our study reveals the limitations of some of the best known master equations
available in the literature, in the context of a spin bath. In general,
perturbative approaches such as low-order NZ\ and TCL do well at short times
(on a time scale set by the system-bath coupling constant) and fare very
poorly at long times. These approximations are also very sensitive to
temperature and do better in the high temperature limit. The PM\ does not do
as well as TCL4 at short times but has the distinct advantage of retaining a
qualitatively correct character for long times. This conclusion depends
heavily on the proper choice of the memory kernel; indeed, when the memory
kernel is not optimally chosen the PM can yield solutions which are not as
satisfactory \cite{ManPet06}.

\acknowledgments
O.O. and H.K. were supported in part by NSF Grant No. CCF-0524822, and H.K 
was also supported in part by NSF Grant No. CCF-0448658. D.A.L. was
supported by NSF Grant No. CCF-0523675.

\begin{widetext}

\appendix

\section{Bath correlation functions}

\label{app:A}

Here we show how to calculate the bath correlation functions used in our
simulations. The $k^{\mathrm{th}}$ order bath correlation function is
defined as 
\begin{equation*}
Q_{k}=\mathrm{Tr}\{B^{k}\rho _{B}\},
\end{equation*}%
where $B$ and $\rho _{B}$ were given in Eqs. (\ref{Bcomp})\ and (\ref%
{eq:rhoB0}), respectively. 
This yields:%
\begin{eqnarray*}
Q_{k} &=&\mathrm{Tr}\{(\sum_{n}g_{n}\sigma _{n}^{z}-\theta I_{B})^{k}\sum_{l}%
\frac{\exp (-\beta E_{l})}{Z}|l\rangle \langle l|\} \\
&=&\sum_{l}\frac{\exp (-\beta E_{l})}{Z}\langle l|(\sum_{n}g_{n}\sigma
_{n}^{z}-\theta I_{B})^{k}|l\rangle \\
&=&\sum_{l,l^{\prime },...,l^{\prime \prime \prime }}\frac{\exp
(-\beta E_{l})}{Z}\langle l|(\sum_{n}g_{n}\sigma _{n}^{z}-\theta
I_{B})|l^{\prime }\rangle \langle l^{\prime }|(\sum_{n^{\prime
}}g_{n^{\prime }}\sigma _{n^{\prime }}^{z}-\theta I_{B})|l^{\prime
\prime }\rangle \langle l^{\prime \prime }|\cdots |l^{\prime
\prime \prime }\rangle \langle l^{\prime \prime \prime
}|(\sum_{n^{\prime \prime \prime}}g_{n^{\prime
\prime \prime}}\sigma _{n^{\prime \prime \prime}}^{z}-\theta I_{B})|l\rangle \\
&=&\sum_{l,l^{\prime },...,l^{\prime \prime \prime }}\frac{%
\exp (-\beta E_{l})}{Z}(\sum_{n}g_{n}\langle l|\sigma
_{n}^{z}|l^{\prime }\rangle -\theta )\delta _{ll^{\prime
}}(\sum_{n^{\prime }}g_{n^{\prime }}\langle l^{\prime }|\sigma
_{n^{\prime }}^{z}|l^{\prime \prime }\rangle -\theta )\delta
_{l^{\prime }l^{\prime \prime }}\cdots (\sum_{n^{\prime \prime
\prime }}g_{n^{\prime \prime \prime}}\langle l^{\prime \prime
\prime }|\sigma _{n^{\prime \prime \prime }}^{z}|l\rangle -\theta
)\delta
_{l^{\prime \prime \prime }l} \\
&=&\sum_{l}\frac{\exp (-\beta E_{l})}{Z}(\sum_{n}g_{n}\langle
l|\sigma _{n}^{z}|l\rangle -\theta )(\sum_{n^{\prime
}}g_{n^{\prime }}\langle l|\sigma _{n^{\prime }}^{z}|l\rangle
-\theta )\cdots (\sum_{n^{\prime \prime \prime }}g_{n^{\prime
\prime \prime }}\langle l|\sigma _{n^{\prime \prime \prime
}}^{z}|l\rangle
-\theta ) \\
&=&\sum_{l}\frac{\exp (-\beta E_{l})}{Z}(\sum_{n}g_{n}\langle
l|\sigma _{n}^{z}|l\rangle -\theta )^{k},
\end{eqnarray*}

or
\begin{equation}
Q_{k}=\frac{1}{Z}\sum_{l}(\tilde{E}_{l})^{k}\exp (-\beta E_{l}),
\end{equation}%
where $Z=\sum_{l}\exp (-\beta E_{l})$ and the expressions for $E_{l}$ and $%
\tilde{E}_{l}$ were given in Eqs. (\ref{eq:El})\ and (\ref{eq:Eitilde}),
respectively.

The above formulas are useful when the energy levels $E_{l}$ and $\tilde{E}%
_{l}$ are highly degenerate, which is the case for example when $g_{n}\equiv
g$ and $\Omega _{n}\equiv \Omega $ for all $n$. For a general choice of
these parameters, it is computationally more efficient to consider $\theta $
in the form (\ref{eq:theta}) and the initial bath density matrix in the form
(\ref{eq_rho_B_inter}). For example, the second order bath correlation
function is
  \begin{eqnarray}
Q_{2} &=&\mathrm{Tr}\{(\sum_{m=1}^{N}g_{m}\sigma _{m}^{z}-\theta
I)(\sum_{n=1}^{N}g_{n}\sigma _{n}^{z}-\theta I)\rho _{B}\}  \notag \\
&=&\mathrm{Tr}\{\sum_{n,m=1}^{N}g_{n}g_{m}\sigma _{n}^{z}\sigma _{m}^{z}\rho
_{B}\} -2\theta \underbrace{\mathrm{Tr}\{\sum_{n=1}^{N}g_{n}\sigma
_{n}^{z}\rho _{B}\}}_{\theta }+\theta ^{2}  \notag \\
&=&\mathrm{Tr}\{\sum_{n,m=1}^{N}g_{n}g_{m}\sigma _{n}^{z}\sigma
_{m}^{z}\bigotimes\limits_{n=1}^{N}\frac{1}{2}(I+\beta _{n}\sigma
_{n}^{z})\}-\theta ^{2}  \notag \\
&=&\sum_{n\neq m}^{N}\mathrm{Tr}\{g_{m}\frac{1}{2}(\sigma _{m}^{z}+\beta
_{m}I)\}\mathrm{Tr}\{g_{n}\frac{1}{2}(\sigma _{n}^{z}+\beta
_{n}I)\}\prod\limits_{j\neq m,n}\mathrm{Tr}\{\frac{1}{2}(I+\beta _{j}\sigma
_{j}^{z})\}+\mathrm{Tr}\{\sum_{n=1}^{N}g_{n}^{2}\rho _{B}\}-\theta ^{2}
\notag \\
&=&\underbrace{\sum_{n,m=1}^{N}g_{m}\beta _{m}g_{n}\beta _{n}}_{\theta
^{2}}-\sum_{n=1}^{N}g_{n}^{2}\beta _{n}^{2}+\sum_{n=1}^{N}g_{n}^{2}-\theta
^{2}  \notag \\
&=&\sum_{n=1}^{N}g_{n}^{2}(1-\beta _{n}^{2}).
  \end{eqnarray}%
Using the identity $1-\tanh ^{2}(-x/2)=2/(1+\cosh x)$, this correlation
function can be expressed in terms of the bath spectral density function
[Eq. (\ref{eq:J})] as follows:%
\begin{eqnarray*}
Q_{2} &=&\sum_{n=1}^{N}g_{n}^{2}(1-\beta _{n}^{2}) \\
&=&\int_{-\infty }^{\infty }\delta (\Omega -\Omega _{n})|g_{n}|^{2}(1-\tanh
^{2}(-\frac{\Omega }{2kT}))\mathrm{d}\Omega \\
&=&\int_{-\infty }^{\infty }\frac{2J(\Omega )\mathrm{d}\Omega }{1+\cosh (%
\frac{\Omega }{kT})}.
\end{eqnarray*}

Higher order correlation functions are computed analogously.

\section{Cumulants for the NZ and TCL master equations}

\label{app:B}

We calculate the explicit expressions for the cumulants appearing in Eq. (%
\ref{eq:cumulants}), needed to find the NZ and TCL perturbation expansions
up to fourth order.

Second order:

\begin{eqnarray}
\langle \mathcal{L}^{2}\rangle \rho  &=&-\mathrm{Tr}_{B}\{[H_{I},[H_{I},\rho
]]\}\otimes \rho _{B}  \notag \\
&=&-\mathrm{Tr}_{B}\{H_{I}^{2}\rho -2H_{I}\rho H_{I}+\rho H_{I}^{2}\}\otimes
\rho _{B}  \notag \\
&=&-2Q_{2}(\rho _{S}-\sigma _{z}\rho _{S}\sigma _{z})\otimes \rho _{B}
\notag \\
&\equiv &\rho ^{\prime },
\end{eqnarray}%
\begin{eqnarray}
\langle \mathcal{L}^{2}\rangle ^{2}\rho  &=&\mathcal{P}\mathcal{L}^{2}%
\mathcal{P}\mathcal{P}\mathcal{L}^{2}\mathcal{P}\rho   \notag \\
&=&\mathcal{P}\mathcal{L}^{2}\mathcal{P}\rho ^{\prime }  \notag \\
&=&-2Q_{2}(\rho _{S}^{\prime }-\sigma _{z}\rho _{S}^{\prime }\sigma
_{z})\otimes \rho _{B},  \notag
\end{eqnarray}%
where $\rho _{S}^{\prime }=\mathrm{Tr}_{B}{\rho }^{\prime
}=-2Q_{2}(\rho _{S}-\sigma _{z}\rho _{S}\sigma _{z})$. Therefore
\begin{eqnarray}
\langle \mathcal{L}^{2}\rangle ^{2}\rho  &=&-2Q_{2}\{(-2Q_{2}(\rho
_{S}-\sigma _{z}\rho _{S}\sigma _{z}))-\sigma _{z}(-2Q_{2}(\rho _{S}-\sigma
_{z}\rho _{S}\sigma _{z}))\sigma _{z}\}\otimes \rho _{B}  \notag \\
&=&8Q_{2}^{2}(\rho _{S}-\sigma _{z}\rho _{S}\sigma _{z})\otimes \rho _{B}.
\end{eqnarray}%
Third order:
\begin{eqnarray}
\langle \mathcal{L}^{3}\rangle \rho  &=&i\mathrm{Tr}_{B}%
\{[H_{I},[H_{I},[H_{I},\rho ]]]\}\otimes \rho _{B}  \notag \\
&=&i\mathrm{Tr}_{B}\{H_{I}^{3}\rho -3H_{I}^{2}\rho H_{I}+3H_{I}\rho
H_{I}^{2}-\rho H_{I}^{3}\}\otimes \rho _{B}  \notag \\
&=&4iQ_{3}(\sigma _{z}\rho _{S}-\rho _{S}\sigma _{z})\otimes \rho _{B}.
\end{eqnarray}%
Fourth order:
\begin{eqnarray}
\langle \mathcal{L}^{4}\rangle \rho  &=&\mathrm{Tr}_{B}%
\{[H_{I},[H_{I},[H_{I},[H_{I},\rho ]]]]\}\otimes \rho _{B}  \notag \\
&=&\mathrm{Tr}_{B}\{H_{I}^{4}\rho -4H_{I}^{3}\rho H_{I}+6H_{I}^{2}\rho
H_{I}^{2}-4H_{I}\rho H_{I}^{3}+\rho H_{I}^{4}\}\otimes \rho _{B}  \notag \\
&=&8Q_{4}(\rho _{S}-\sigma _{z}\rho _{S}\sigma _{z})\otimes \rho _{B}.
\end{eqnarray}

\end{widetext}

%\bibliographystyle{/home/Daniel/revtex/prsty}
%\bibliography{/home/Daniel/articles/bib}

\end{document}